\documentclass[%
 reprint,
 amsmath,amssymb,
 aps,
floatfix,
]{revtex4-2}

\usepackage{graphicx}
\usepackage{dcolumn}
\usepackage{xcolor}
\usepackage{subfigure}
\usepackage{amsmath}
\usepackage{amssymb} 
\usepackage{mathrsfs}
\usepackage{color}
\usepackage{multirow}
\usepackage[normalem]{ulem}
\usepackage{bm}
\usepackage{array}
\usepackage{placeins}

\begin{document}


\title{Gravity-driven flux of particles through apertures}

\author{Ram Sudhir Sharma,$^{1,*}$  
Alexandre Leonelli,$^{1}$
Kevin Zhao,$^{2}$  
Eckart Meiburg,$^{1}$
and Alban Sauret$^{3,4}$
}\email{ramsharma@ucsb.edu}
\affiliation{$^{1}$ Department of Mechanical Engineering, University of California, Santa Barbara, CA 93106, USA\\
$^{2}$ Department of Physics, University of California, Santa Barbara, CA 93106, USA \\
$^{3}$ Department of Mechanical Engineering, University of Maryland, College Park, MD 20742, USA \\
$^{4}$ Department of Chemical and Biomolecular Engineering, University of Maryland, College Park, Maryland 20742, USA
}

\date{\today}
\begin{abstract}
The gravity-driven discharge of granular material through an aperture is a fundamental problem in granular physics and is classically described by empirical laws with different fitting parameters. In this Letter, we disentangle the mass flux into distinct velocity and packing contributions by combining three-dimensional experiments and simulations. 
We define a dimensionless flux ratio that captures confinement-driven deviations from a free-fall limit, which is recovered when the aperture is large compared to the grain size. 
For spherical cohesionless grains, the deviations from the free-fall limit are captured by a single exponential correction factor over a characteristic length scale of $\sim$ 10-15 grain diameters. This is shown to be the scale over which the packing structure is modified due to the boundary. Building on the $\sqrt{gD}$ exit–velocity scaling, we propose a kinematic framework that explains the universality of granular discharge beyond empirical descriptions.
\end{abstract}

\maketitle

Transport through bottlenecks is a ubiquitous physical process. In the simplest case of a Newtonian liquid, the discharge reflects the interplay between hydrostatic pressure and the aperture area, as captured by Torricelli’s law \cite{ferrand2016wetting}. For more complex materials through bottlenecks such as particulate suspensions \cite{marin2025clogging, dressaire2017clogging}, polymer solution extrusion \cite{larson1999structure}, foam drainage \cite{bertho2006dense, dollet2010local, schott2023three}, or crowds moving through constrictions \cite{garcimartin2015flow, garcimartin2016flow}, the flow rate depends sensitively on how the microstructure of the flowing material couples to the geometry of the opening. Complexities in viscosity, elasticity, and yield stress can modify transport \cite{bonn2017yield, teoman2022discharge} and switch the discharge from dripping to jetting \cite{clanet1999transition, utada2007dripping, waitukaitis2011droplet}. \smallskip

Gravity-driven discharge of dry particles through apertures displays behavior that is deceptively similar to fluids yet governed by entirely different physics. Many particulate discharges resemble liquid jets \cite{prado2013incompressible, boudet2007granular}. However, granular discharge shows two important features that set it apart from ordinary fluids. First, once the flow is established, the mass flux $Q$ through a basal aperture is constant for most of the drainage process, unlike an ordinary fluid \cite{nedderman1982flow, jaeger1996granular}. Second, when the ratio $D/d$ between the aperture diameter $D$ and the grain diameter $d$ becomes small, arch formation and intermittent clogging are observed \cite{zuriguel2005jamming, zuriguel2014invited}. \smallskip

\begin{figure}
\centering
\includegraphics[width =0.95\columnwidth]{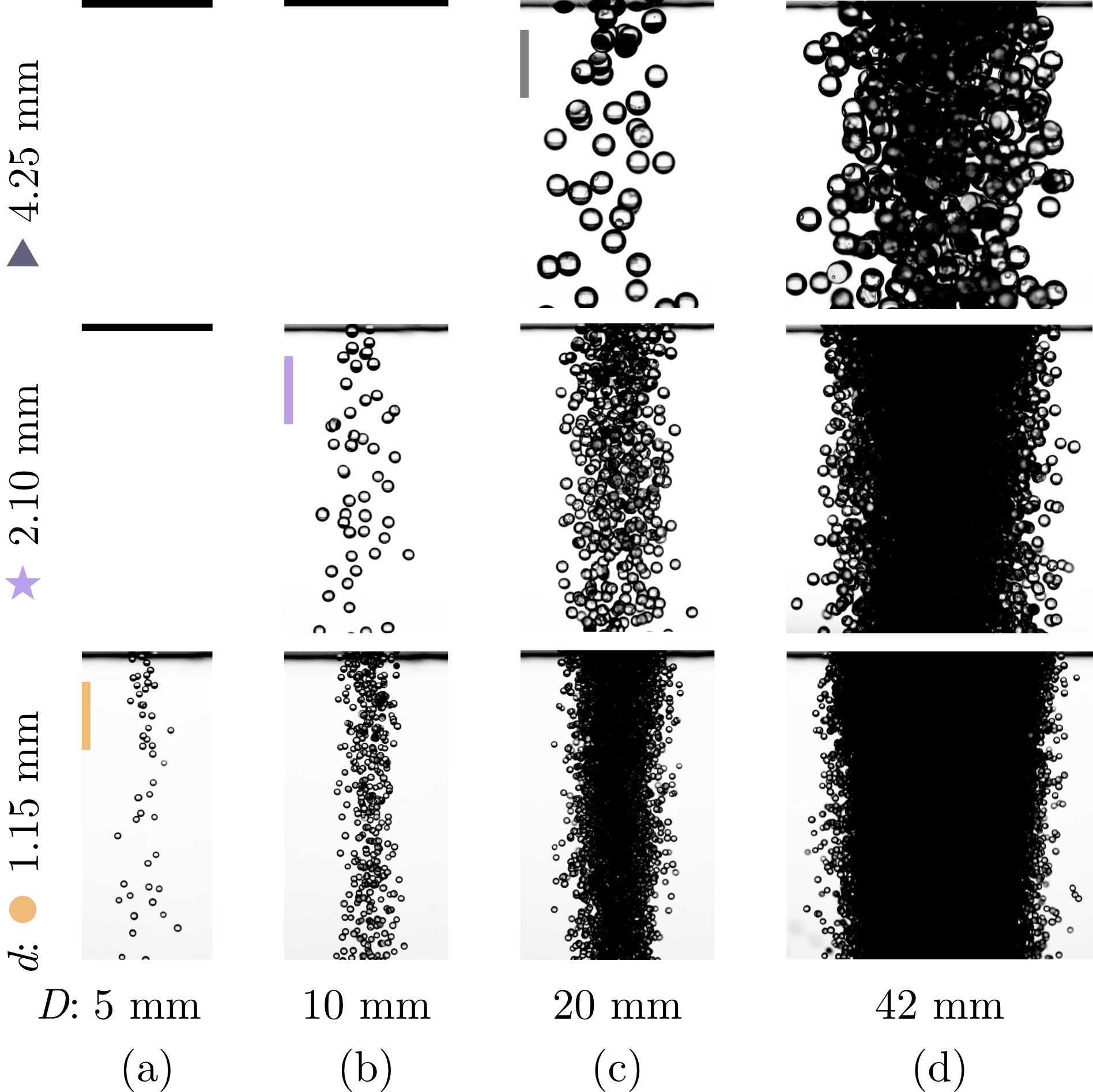}
\caption{Examples of granular discharge for particles of different diameters $d$ through circular apertures of diameter $D$: (a) $5$ mm, (b) $10$ mm, (c) $20$ mm, (d) $42$ mm. Rows correspond to $d=4.25, \, 2.10, \, 1.15$ mm from top to bottom. Scale bars are 1 cm. For larger $D/d$, denser flows are observed. No flows are observed for $D/d \lesssim 3$.}
\label{fig:Fig1_Imaging}
\end{figure}

While this constant discharge of dry particles has been studied for more than a hundred and fifty years \cite{tighe2007pressure}, empirical relations for the constant flux $Q$ in terms of aperture size $D$, grain size $d$, bulk density $\rho_b$, and gravity are still used \cite{fowler1959flow,beverloo1961flow}. One of the most widely used is Beverloo's relation \cite{beverloo1961flow}: 
\begin{equation}
Q = C \rho_b \sqrt{g} \left(D - kd\right)^{5/2}.
\label{eq:eq_Beverloo}
\end{equation} 
This expression relies on two fitting parameters: $C$, accounting for the shape of the opening (usually $\simeq 0.58$ for a circular aperture \cite{nedderman1982flow}), and $k$ ($\simeq 1.5$ for spherical particles \cite{beverloo1961flow,nedderman1982flow}), which defines an effective aperture size $D-kd$. \citet{mankoc2007flow} performed experiments over a wider range of apertures and developed a relation by quantifying the deviations between their measurements and Beverloo's relation \eqref{eq:eq_Beverloo} with an exponential correction proposed. Additional experiments were later performed by \citet{benyamine2014discharge}, and an expression of the flux was provided as:
\begin{equation} 
Q = C' \, A \, \rho_b \, \sqrt{gD} \, \left[1 - \alpha_1 \, e^{-\alpha_2(D/d)}\right], 
\label{eq:eq_Benyamine} 
\end{equation} where $A$ is the area of the aperture. This expression removes the dependence on the parameter $k$ but introduces two additional fitting parameters to Eq. \eqref{eq:eq_Beverloo} besides $C' \simeq 0.75$, $\alpha_1 \simeq 0.96$ and $\alpha_2 \simeq 0.09$  \cite{benyamine2014discharge}.  While this expression lacks a clear physical interpretation, the exponential term broadly captures dilation as observed on the structure of discharge (Fig.~\ref{fig:Fig1_Imaging}). In this Letter, we revisit the problem of steady flux in particulate discharge and propose a framework rooted in free-fall kinematics. To this end, a large number of experiments and Discrete Element Method (DEM) simulations are carried out for a range of $d$ and $D$. \smallskip

\begin{figure}
\centering
\includegraphics[width =\columnwidth]{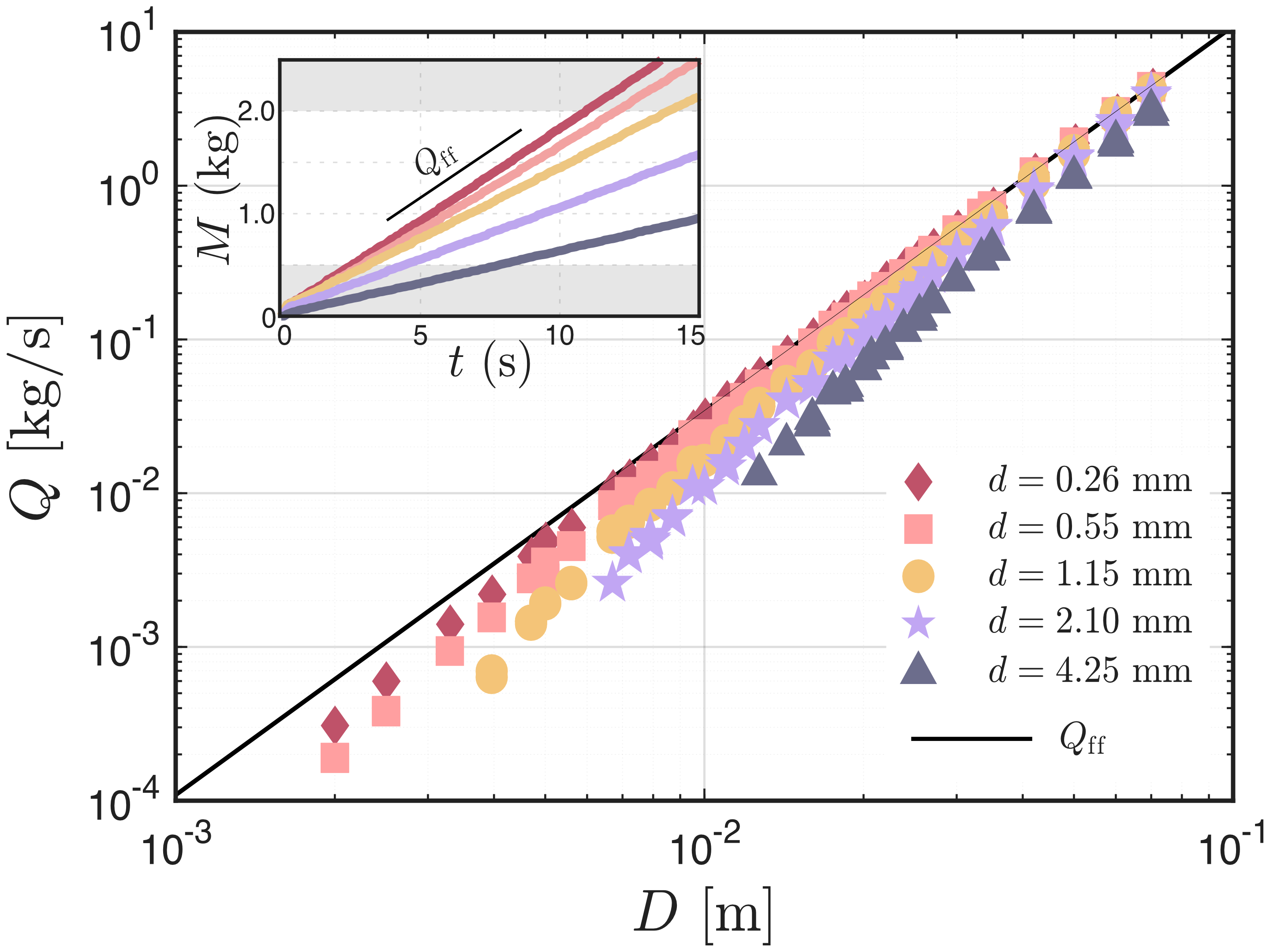}
\caption{Measured mass flux $Q$ from discharge experiments for a range of $D$ and $d$. The solid line shows the model free-fall flux given by Eq. \eqref{eq:eq_freeFall_flux}, $Q_{\rm ff} = A \, \rho_g \sqrt{g\, D} \, \phi_{\rm ff}$, independent of $d$. Inset: Mass–time data $M(t)$ for $D=20$ mm [Fig.~\ref{fig:Fig1_Imaging}(c)] across all grain sizes in the region of interest. The slope of each curve yields $Q$. For $D \gg d, \, Q_{\rm ff}$ predicts the flux.}
\label{fig:Fig2_Cohesionless_Data}
\end{figure}

For our experiments, a cylindrical flat-bottomed silo is initially filled with $3\,$-$7$ kg of particles, using a narrow funnel such that the particles are rained in randomly. 
The aperture at the base is then opened and the falling mass is drained onto a container on a weighing scale (OHAUS EX602 with $\pm 0.01$ g accuracy and $10\,{\rm Hz}$  acquisition), leading to the mass flow rate $Q = \Delta M / \Delta t$. 
The particles used are glass spheres (Sigmund Lindner, GmbH) of mean diameter $d = 0.26, \, 0.55, \, 1.15, \, 2.10 \,\, \& \, 4.25$ mm and density $\rho_g\approx 2.5\, \text{g.cm}^{-3}$. We use circular apertures with diameters $2\, \text{mm}\leq D \leq 70 \, \text{mm}$ placed at the base of the cylinder. Experiments are conducted in three cylindrical silos: one made of stainless steel (inner diameter $D_{\rm cyl} = 10.2$ cm), and two PMMA cylinders ($D_{\rm cyl} = 10$ and $20$ cm). In all cases, we ensure $D_{\rm cyl} \gg D$, so for this problem, only two length-scales are \textit{a priori} relevant: $D$ and $d$. At least three independent trials are conducted for each combination of $D$ and $d$. In all cases, prior to starting, $\phi_{\rm cyl} \approx 0.60$ and $\rho_b = \rho_g \phi_{\rm cyl} \approx 1.5 \, \text{g.cm}^{-3}$. Details on particle properties (size, density, packing at rest) and apertures (sizes, materials, thicknesses), as well as the experimental procedure, are provided in Supplementary Materials S.I. \smallskip

Snapshots of some discharges having just been through the aperture are shown in Fig.~\ref{fig:Fig1_Imaging}. The corresponding videos are shown in Supplementary Materials. Example measurements of mass discharge for each particle size through an aperture $D=20$ mm are shown in the inset of Fig.~\ref{fig:Fig2_Cohesionless_Data}, corresponding to Fig.~\ref{fig:Fig1_Imaging}(c). Measurements from all flux experiments are summarized in Fig. \ref{fig:Fig2_Cohesionless_Data}. Altogether, these experiments show that when the aperture is much larger than the particles passing through it, \textit{i.e.}, $D \gg d$, the observed discharge and the measured flux approach a dense asymptotic value. We first describe a model free fall flux $Q_{\rm ff}$ for this case, \textit{i.e.}, drainage driven by gravity and independent of grain size. The general description of mass flux through an aperture is:
\begin{equation}
    Q = \int_{\mathrm{A}} \vec{q} \cdot {\rm d}\vec{A} = A \,  \rho_g \, \langle u_z \, \phi \rangle_{A},
    \label{eq:eq_massFlux_general}
\end{equation} where $\langle u_z \, \phi\rangle_A$ is the product of the vertical velocity and the packing fraction averaged over the area $A=\pi\,D^2/4$ of the aperture at the silo base $z=0$. Very close to the opening, the jet of grains has the same cross-sectional shape as the aperture through which they will fall, and consequently, $A$ is unmodified in the free-fall idealization. Since measuring the velocity and packing distributions in 3D is complex \cite{van1974density}, studies have relied on numerical simulations \cite{maza2013velocity, zhou2015discharge} or 2D analogues with a single grain layer flowing through a slit \cite{janda2012flow, gella2017role}. We present simulations using the open source DEM software LIGGGHTS in S.III. \cite{liggghts}, where the geometry of our experiments are replicated and standard micro-mechanical parameters are used \cite{dumont_micro_params_2020}. While the distributions of $z$-velocity and packing cannot be separated in our experiments, this is confirmed using simulations. We calculate the covariance as $\langle u_z \, \phi \rangle_A - \langle u_z \rangle_A \langle \phi\rangle_A$, and find it to contribute an error of $\approx 1.5\%$ for small $D/d$, and $< 0.01\%$ for $D \geq  20d$. The decomposition $\langle u_z \, \phi \rangle_A = \langle u_z \rangle_A \langle \phi \rangle_A$ is also a common assumption in prior works \cite{maza2013velocity, zhou2015discharge, janda2012flow, gella2017role}. The distributions of velocity and packing at the aperture are obtained using an Eulerian framework \cite{kempe_level_set_2013} and discussed in S.IV. and S.V., respectively. Here, we summarize the arguments to propose a simplified form of the flux in free-fall, $Q_{\rm ff}$. \smallskip

\begin{figure}
\centering
\includegraphics[width =0.97\columnwidth]{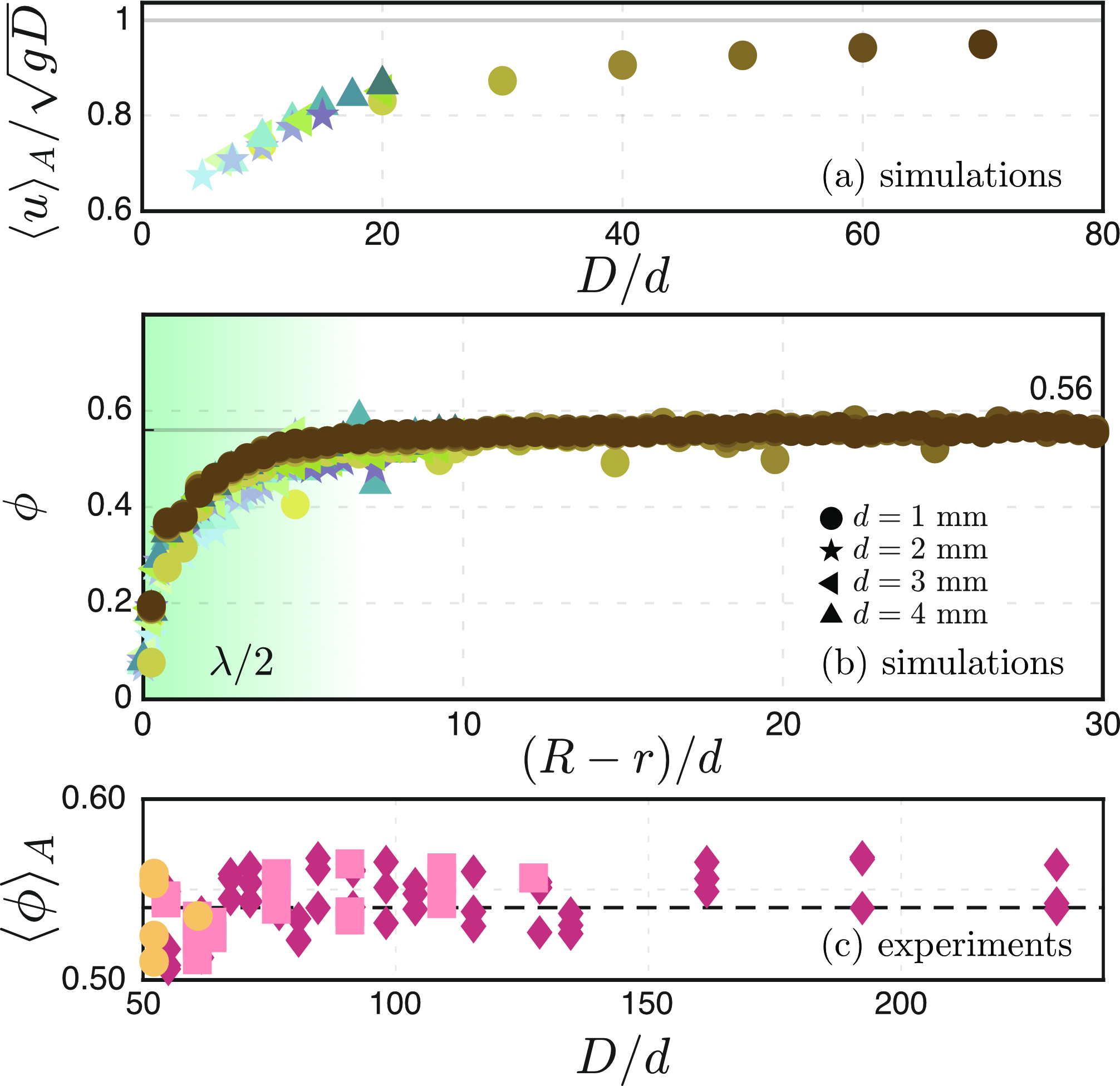}
\caption{(a) DEM simulations of the mean vertical velocity at the aperture, $\langle u_z \rangle_A / \sqrt{gD}$, for various $d$. Values remain between 0.8 and 1, approaching 1 for $D \gg d$. 
(b) Packing fraction from all simulations collapses in wall-normal coordinates when rescaled by particle size ($R = D/2$). Far from walls (simulations): $\phi_{\rm ff} = 0.56$.
(c) Area-averaged packing fraction, $\langle \phi \rangle_A = Q / A\rho_g\sqrt{gD}$, from experiments with $D \geq 50d$. Mean value of (experiments) $\phi_{\rm ff} \simeq 0.54$ (dashed line).}
\label{fig:Fig3_Velocity_Packing}
\end{figure}

In free-fall, the boundary of the aperture is assumed to minimally affect the flux. Stresses in the column above the aperture are neglected \cite{aguirre2010pressure}, and grains are taken to accelerate under gravity over a characteristic distance $L$, giving $\langle u_z \rangle_A = \sqrt{2gL}$. Prior studies indicate that the free-fall region is localized over a distance $D$ above the orifice \cite{zuriguel2005jamming,rubio2015disentangling}. While often described as a ``free-fall arch," no singular structure exists. Instead, \citet{rubio2015disentangling} identified a kinematic boundary at a height $D/2$, where kinetic stresses peak and contacts rapidly vanish. This motivates $L = D/2$, and a characteristic free–fall velocity scale of $\sqrt{2gL} = \sqrt{gD}$, which is a classically expected value. In the 2D analogue to the geometry here, a number of studies found that the velocity of particles at the aperture broadly scale with $\sqrt{gD}$, with a slightly larger value in the center of the aperture \cite{janda2012flow, dorbolo2013influence, gella2017role}. We propose that the consequence of the boundary identified by \citet{rubio2015disentangling} must be observed on the mean velocity of particles as they pass through the aperture in free-fall, rather than the centerline: \begin{equation}
\langle u_z \rangle_{A,\,{\rm ff}} \simeq \sqrt{gD}.
\end{equation} DEM simulations confirm this expectation: $\langle u_z \rangle_A / \sqrt{gD}$ remains above $0.7$ for all $D/d$, and approaches unity as $D \gg d$ as shown in Fig.~\ref{fig:Fig3_Velocity_Packing}(a). Moreover, when particle velocities are averaged in time and over the angular coordinate, the radial velocity profiles from simulations converge toward a simple form for $D \gg d$. 
Following an argument presented for the 2D analogue by \citet{janda2012flow}, we propose the converging asymptotic form for $D/d \gg 1$ in 3D to be [Fig.~S2(e)]
\begin{equation} \label{eq:velo}
u_{z,\,{\rm ff}}(r) = \frac{5}{4}\sqrt{gD}\,\left[ 1 - \Big( \frac{2r}{D} \Big)^2 \right]^{1/4}.
\end{equation}
The exponent $1/4$ comes from the free–fall height and the centerline velocity, $5/4 \sqrt{gD}$, is fixed by the condition $\langle u_z\rangle_A = \sqrt{gD}$ (see End Matter and Sec.~S.IV). 
Altogether, these results establish that the velocity contribution to the free-fall flux is primarily set by gravity and aperture size, independent of grain size. \smallskip

In constructing $Q_{\rm ff}$, we would like to use a measure of packing, $\phi_{\rm ff}$, at the onset of contact breakdown. We expect $\phi_{\rm ff} < \phi_{\rm cyl}$, but no theoretical prediction for such a free-fall packing exists. We therefore estimate $\phi_{\rm ff}$ from experiments using mass conservation at the aperture. For $D \gg d$, $\langle u_z \rangle_A = \sqrt{gD}$ is a reasonable estimate (Fig.~\ref{fig:Fig3_Velocity_Packing}a). Consequently, for $D > 50d$, we determine $\langle \phi \rangle_A = Q/(A \rho_g \sqrt{gD})$, as shown in Fig.~\ref{fig:Fig3_Velocity_Packing}(c). Across grain sizes, this yields $\phi_{\rm ff} \simeq 0.54 $ (RMSE: $0.02$). Our DEM simulations with monodispersed spheres converge to a very comparable asymptotic value of $\phi_{\rm ff} \simeq 0.56$, with a clear plug-like distribution when viewed in wall-normal coordinates [Fig.~\ref{fig:Fig3_Velocity_Packing}(b), described in End Matter]. Notably, our estimates of $\phi_{\rm ff}$ closely match the volume fraction measured from incompressible granular jets \cite{prado2013incompressible}. Classic experiments by \citet{onoda1990random} determined a lower limit to the random loose packing of cohesionless spheres, $\phi_{\rm RLP} (g \rightarrow 0) \simeq 0.555 \pm 0.005$, corresponding to the onset of rigidity percolation, in close agreement with our experimental and numerical results. This suggests that the asymptotic free-fall state coincides with the threshold at which network rigidity is lost, \textit{i.e.}, the limit at which grains can begin to dilate freely. Thus, $\phi_{\rm RLP} (g \rightarrow 0)$ provides a natural reference point for $\phi_{\rm ff}$. Fig. S4 shows $\langle \phi \rangle_A \rightarrow \phi_{\rm ff}$ for $D\gg d$. Additional details on packing are provided in S.V. Unlike the velocity scale, which is broadly universal, the packing is sensitive to confinement and boundary effects, and thus plays the central role in setting corrections from an idealized flux, in particular when $D \sim d$. \smallskip

Altogether, we write an idealized free-fall flux as \begin{equation}
    Q_{\rm ff} = A \, \rho_g \sqrt{g\, D} \, \phi_{\rm ff}.
    \label{eq:eq_freeFall_flux}
\end{equation} This equation is plotted as a solid line alongside the experimental data in Fig. \ref{fig:Fig2_Cohesionless_Data}. Note that this expression was derived in the limit $D/d \gg 1$, and depends on the free-fall packing of the system. Our experiments show that when $D \gg d$, Eq.~\eqref{eq:eq_freeFall_flux} successfully predicts the flux. \smallskip

\begin{figure}
\centering
\includegraphics[width =\columnwidth]{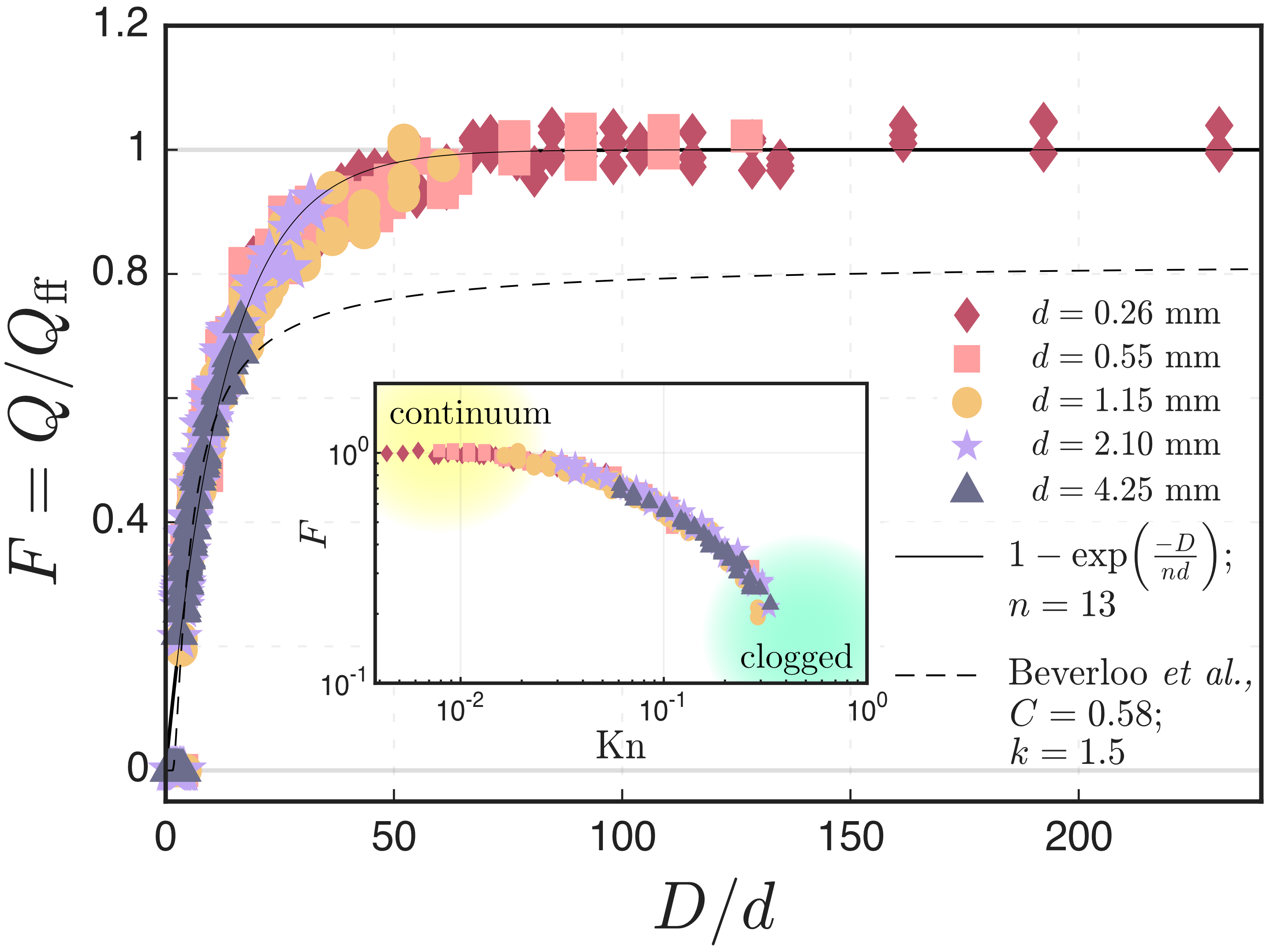}
\caption{Normalized flux $F = Q / Q_{\rm ff}$ vs relative aperture size $D/d$ for all experiments. Data collapse onto a single curve, described by $F = 1 -  {\rm exp}(-D/nd)$, with $n=13$ (solid line) [RMSE: 0.04]. Beverloo’s relation \eqref{eq:eq_Beverloo}, with $C =0.58$, $k = 1.5$, \cite{nedderman1982flow}
and $\phi_b = 0.6$, is shown normalized by $Q_{\rm ff}$ (dashed line). Inset: Evolution of $F$ with a Knudsen number, $\mathrm{Kn} = (D/d)^{-1}$.}
\label{fig:Fig4_Froude_Dd}
\end{figure}

In Fig. \ref{fig:Fig4_Froude_Dd}, we compare $Q_{\rm ff}$ with the experimental measurements of $Q$ for the range of tested $D/d$. Data from all the experiments conducted collapse onto a single trend. Indeed, for $D \gg d$, the ratio of fluxes approaches 1, as $Q \simeq Q_{\rm ff}$ and gravity dominates. When $D \rightarrow d$, the measured flux rapidly approaches $0$, as the effects of confinement become important. We denote the ratio as a dimensionless measure of flux: \begin{equation}
F = \frac{Q}{Q_{\rm ff}} \equiv 
\frac{\langle \phi \, u_z\rangle_A}{\phi_{\rm ff} \, \sqrt{g\, D}}.
\label{eq:eq_Froude}
\end{equation} Confinement, \textit{i.e.}, deviations from $F \simeq 1$, are shown to depend on the relative aperture size $D/d$. When decomposed using simulations (S.III.), both velocity and packing distributions are shown to asymptote towards the free-fall case when $D \gg d$. For velocity, a small correction to the free fall scale is needed when $D\sim d$ (S.IV.). The packing profiles show that when $D \gg d$, the plug-like region  $\phi_{\rm ff}$ dominates the overall packing.
The individual contributions of the velocity and the packing to the overall confinement effects in the transition regime are presented in End Matter. Here, a one-parameter expression is presented to capture the confinement of flux, combining both distributions and trading their explicit separation for parsimony. \begin{equation}
    F = 1 - \exp\!\left(\frac{-D}{nd}\right)
\end{equation} where $nd \equiv \lambda$ is an effective relaxation length. This form is chosen to reflect the picture established in wall-normal coordinates for packing (S.V.). Modifications to the local mass flux density are strongest near the boundary, but relax exponentially toward a plug in the bulk. Each increment of aperture width, therefore, adds an independent ``layer" of grains that dilute boundary effects, in a manner consistent with a first-order relaxation process. Effects of confinement on the overall velocity are also folded into this correction. Fitting to our data across grain sizes leads $\lambda \simeq 10d$–$15d$, \textit{i.e.} $n \simeq 10-15$ particles (Fig.~\ref{fig:Fig4_Froude_Dd} with $n = 13$), in agreement with the characteristic scale that often separates particle and bulk behavior in granular systems \cite{pouliquen1999scaling, prado2013incompressible}. Thus, $\lambda$ sets the scale of boundary influence. The region of strong dilation is confined to $\sim\lambda/2$ [7–8 grain diameters from the wall; see Fig.~\ref{fig:Fig3_Velocity_Packing}(b)], where the geometric effects of excluded volume ensure a packing deficit near the wall, that recovers to the value $\phi_{\rm ff}$ over a few particle diameters \cite{benenati1962void, jenkins1983theory}. The exponential correction provides a simple, one-parameter expression for the gravity-driven flux, \begin{equation}
    Q = A \, \rho_g \sqrt{g D} \, \phi_{\rm ff} \, \Big( 1 - e^{-D/nd} \Big).
    \label{eq:eq_fluxNew}
\end{equation} The magnitude of $n$ is not expected to be universal, but will depend on particle shape, boundary interactions, or cohesive forces which would all modify the relaxation of the packing and velocity distributions to their asymptotic values. For comparison, Fig.~\ref{fig:Fig4_Froude_Dd} also shows Beverloo’s relation \eqref{eq:eq_Beverloo} with $C = 0.58$, $k=1.5$ \cite{nedderman1982flow}, and bulk packing $\phi_b = 0.60$. While Beverloo captures trends for $D \lesssim 10d$, clear deviations emerge for $D \gg 10d$. \smallskip

For $D/d \lesssim 6$, grains clog stochastically, and for $D/d \lesssim 3$, no flux occurs. The exponential correction presented here only describes the flowing states. Nevertheless, the same structural ingredients, \textit{i.e.}, the free-fall packing $\phi_{\rm ff} \simeq \phi_{\rm RLP}$ and the boundary-layer length $\lambda \simeq nd$ naturally suggest a route to clogging as a rigidity onset \cite{zuriguel2003jamming}. As $D/d$ decreases and the dilated boundary layer occupies a finite fraction of the aperture, a percolating contact network can intermittently support load, producing a finite clog probability. Beyond this threshold, when flows are established, the dimensionless flux ratio $F$ plays the role of a granular Froude number. By analogy to fluid jets, $\mathrm{Fr}^2 = u_o^2 / (gD)$ compares inertia to gravity \cite{milne1996theoretical} where $u_o$ is its initial velocity. Here $F$ is augmented by the factor $\langle \phi \rangle_A / \phi_{\rm ff}$, reflecting the distribution of mass. In the inset of Fig.~\ref{fig:Fig4_Froude_Dd}, $F$ is plotted as a function of a Knudsen number. $\mathrm{Kn}$ compares $D$, the macroscopic scale with $\ell$, a mean free path. In molecular fluids, $\mathrm{Kn} = \ell/D \lesssim 10^{-2}$ describes a region where a notion of continuum can be utilized \cite{chambre2017flow}. Remarkably, if $\ell \sim \mathcal{O} (d)$ \cite{sone2007molecular, jenkins1983theory}, a similar transition is observed with grains as $F \rightarrow 1$. \smallskip

In this Letter, we have revisited the mass flux of particles through apertures and demonstrated a minimal expression. Specifically, the flux is set by three non-trivial ingredients: (i) an average velocity scale that approaches the classical form containing gravity and the aperture for $D \gg d$, (ii) a free-fall packing far from edges, or when $D \gg d$, and (iii) the confinement corrections that arise when $D$ is comparable to $d$. This framework can be extended to less ideal conditions — sharp-edged apertures, rough or angular grains, and cohesive interactions. Each case introduces additional heterogeneity through the relevant length scales. While the Froude-like framework remains broadly robust, these cases highlight how boundary conditions and material properties shape real granular discharge beyond the idealized geometry presented here. Our results reveal why hourglasses have been so reliable for centuries: their constancy reflects the universality of free fall, together with confinement-modulated packing.

\section*{Acknowledgements}
We acknowledge Jacob Winefeld and Jonathan Xue for contributing to preliminary experiments and simulations presented here. This work was supported by the National Science Foundation Particulate and Multiphase Processes program under Grant No. 2533460 and by the U.S. Army Research Office under Grant No. W911NF-23-2-0046.

\bibliographystyle{apsrev4-2}
\bibliography{BEV.bib}

\section*{End Matter}

\subsection{Velocity Profiles ($D\gg d$)}
Following the free-fall model of \citet{janda2012flow} in 2D, we consider in 3D a hemispherical dome of base diameter $D$ above the aperture, from which particles fall from rest. Although no physical dome exists \cite{rubio2015disentangling}, it provides a useful kinematic reference. The free-fall height is then $h(r) = (R^2 - r^2)^{1/2}$, with $R = D/2$ giving \begin{equation}
    u_{z}(r) = B \sqrt{2g\, h(r)} = B \sqrt{2gR} \,\,\, f(r)
    \label{eq:S_velocity_general}
\end{equation} where details of the shape are captured through the function $f(r) = \left[1 - (r/R)^2\right]^{1/4}$. Enforcing $\langle u_z \rangle_A = \sqrt{2gR}$, yields $B = 1/\langle f(r) \rangle_A= 5/4$ such that \begin{equation}
    u_{z,\, \mathrm{ff}} (r) = \frac{5}{4} \, \sqrt{2gR} \, \left[1 - \Big(\frac{r}{R}\Big)^2\right]^{1/4}. 
    \label{eq:velocity_toy}
\end{equation} Simulation profiles [Fig. S2 (e)] approach this free-fall prediction at large $D/d$. The model predicts a centerline velocity $u_c = B \, \sqrt{gD} = (5/4)\sqrt{gD}$ in the limit $D \gg d$. In Fig. S3, we show $u_{\rm c}$ and the corresponding mean velocities $\langle u_z \rangle_A$ over a large range of $D/d$. \smallskip

\subsection{Packing Profiles}

Expressed in wall-normal coordinates $x = R - r$, all packing profiles from the simulation data collapse when normalized by grain size $d$ (Fig. S5 (e)). The transient region extends over $\lambda/2 \simeq 7-8d$, consistent with the scale $nd$ extracted from the flux ratio ($n \approx 10-15$). The relaxation near the wall behaves as a structural analogue of a Knudsen layer. Borrowing the exponential form from gas kinetics \cite{sone2007molecular}, we have \begin{equation}
    \phi(x) = \phi_{\rm ff} \left[1 - {\rm exp} \Big({-\frac{x}{\ell}}\Big)\right],
\end{equation} with $\ell \simeq 1-2\, d$ from simulations [Fig. S5 (e), best fit: $\ell = 1.2\, d$]. Recasting the packing into a radial form, \begin{equation}
    \phi(r) = \phi_{\rm ff} \left[1 - {\rm exp} \Big({\frac{r-R}{\ell}}\Big)\right],
    \label{eq:eq_phi_full_final}
\end{equation} we obtain the aperture averaged form as:\begin{equation}
    \frac{\langle \phi \rangle_A}{\phi_{\rm ff}} = 1 - \frac{4 \ell}{D} + \frac{8\ell^2}{D^2}\left[1 - {\rm exp}\Big(-\frac{D}{2\ell}\Big)\right].
\end{equation} Fits to experiments are also well described by $\ell \simeq 1\, d$ (Fig. S4). \smallskip

This exponential recovery of packing can be connected to wall-loosening or excluded-volume forms known from dense packings \cite{benenati1962void, jenkins1983theory}. Particle centers are geometrically forbidden within one grain diameter of the wall, and the packing fraction recovers exponentially over the correlation length $\ell$ \cite{jenkins1983theory}. The Knudsen-like layer and this excluded-volume relaxation are therefore two views of the same geometry: the former a continuum boundary-layer analogy, the latter its microscopic origin. Both describe the same structural field that defines the confinement length $\ell \simeq d$. \smallskip

\subsection{Combining distributions to $F$ when $D \sim d$}

The dimensionless flux ratio $F$ is \begin{equation}
    F = \frac{\langle u_z \phi\rangle_A}{\sqrt{gD}\, \phi_{\rm ff}} = \Big(\frac{\langle u_z\rangle_A}{\sqrt{gD}}\Big)\Big(\frac{\langle \phi\rangle_A}{\phi_{\rm ff}}\Big).
\end{equation} From our simulations, when $D/d \gtrsim 1$ the ratio of velocities is described by $\langle u_z \rangle_A/\sqrt{gD} = 1 - (D/d)^{\beta}$ (best fit: $\beta=-0.63$). Combining the two contributions yields \begin{equation}
    F = \Big[ 1 - \Big(\frac{D}{d}\Big)^{\beta}\Big] \Big[ 1 - 4\Big(\frac{D}{\ell}\Big)^{-1} + 8\Big(\frac{D}{\ell}\Big)^{-2}\left(1 - e^{-\frac{D}{2\ell}}\right)\Big].
    \label{eq:eq_flux_full_micro}
\end{equation} Equation \eqref{eq:eq_flux_full_micro} reproduces the combined confinement effects: rapid suppression of flux at small $D/d$ from exponential packing relaxation, and a slower algebraic approach to the free-fall limit as $D/d$ increases. The stretched single-parameter exponential of Eq. \eqref{eq:eq_fluxNew} captures the same phenomenon, accurately fitting the collapse from experiments.

\end{document}


\title{Supplementary Materials: \\ Gravity-driven flux of particles through apertures}

\author{Ram Sudhir Sharma,$^{1,*}$  
Alexandre Leonelli,$^{1}$
Kevin Zhao,$^{2}$  
Eckart Meiburg,$^{1}$
and Alban Sauret$^{3,4}$
}\email{ramsharma@ucsb.edu}
\affiliation{$^{1}$ Department of Mechanical Engineering, University of California, Santa Barbara, CA 93106, USA\\
$^{2}$ Department of Physics, University of California, Santa Barbara, CA 93106, USA \\
$^{3}$ Department of Mechanical Engineering, University of Maryland, College Park, MD 20742, USA \\
$^{4}$ Department of Chemical and Biomolecular Engineering, University of Maryland, College Park, Maryland 20742, USA
}

\date{\today}
\maketitle


\section{Details on Experiments}

Experiments were performed in a stainless steel cylinder of inner diameter $D_{\rm cyl} = 10.2$, chosen to minimize electrostatic charging. Experiments with PMMA and glass cylinders of the same diameter confirmed that the wall material had no measurable influence on the flux, provided the cylinder was much wider than the aperture of diameter $D$. For large relative apertures, 
$D / D_{\rm cyl} \gtrsim 0.35$, we used a wider PMMA cylinder ($D_{\rm cyl} = 20$ cm) so that the opening area fraction never exceeded $ \simeq 10\%$ for all our experiments. This ensured that downward advection of the granular column did not significantly contribute to the discharge. \smallskip

The flux was measured from mass-collection data as described in the main article. 
To avoid transient artifacts, we restricted the fits of the data to the first $0.5-2$ kg of discharged material (see inset of Fig. 2, main article). 
This corresponds to early times when the free surface is essentially flat and grain advection is minimal. 
Using later times slightly biases the measure of flux, and thereby $\phi_{\rm ff}$, as we confirmed by comparison with DEM simulations. For very small fluxes, a time cutoff is applied instead, using a fixed $30$ s window of $M(t)$, starting $1$ s after opening the aperture.\smallskip

\begin{figure}
\centering
\includegraphics[width =0.95\columnwidth]{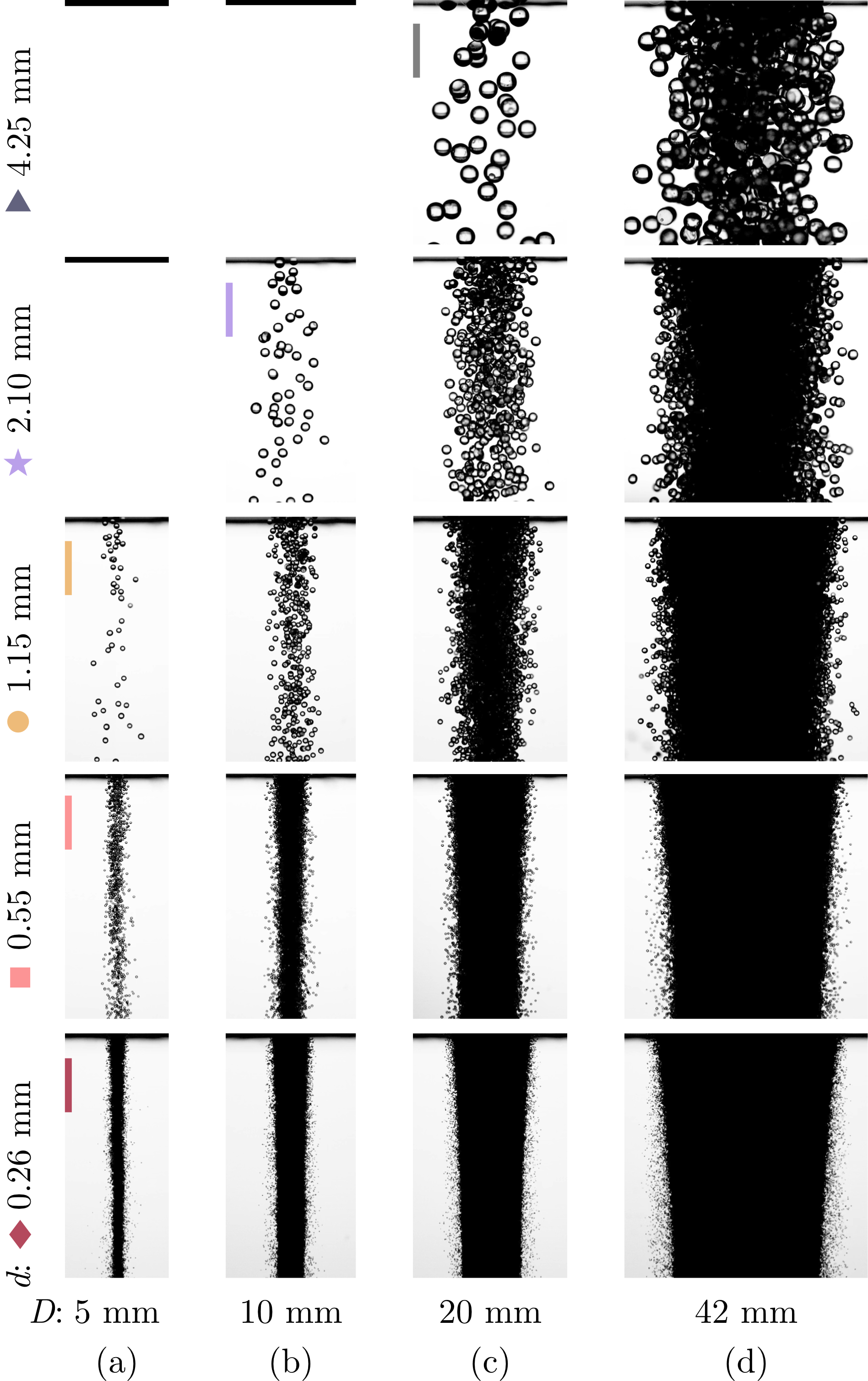}
\caption{Examples of granular discharge for particles of different diameters $d$ through circular apertures of diameter $D$: (a) $5$ mm, (b) $10$ mm, (c) $20$ mm, (d) $42$ mm. Rows correspond to $d=4.25, \, 2.10, \, 1.15, \, 0.55, \, 0.26$ mm from top to bottom. Scale bars are 1 cm. No flows are observed for $D/d \lesssim 3$.}
\label{fig:SFig1_Imaging}
\end{figure}

Glass spheres (Sigmund Lindner GmbH) with mean diameters $d = 0.26, \, 0.55, \, 1.15, \, 2.10$ and $4.25$ mm ($\sim 10\%$ polydispersity) were used, with density 
$2500-2560$ kg.m$^{-3}$. Their static packing inside the cylindrical silo was $\phi_{\rm cyl} = 0.60-0.61$ (Table~\ref{tab:particle_properties}). \smallskip

Circular apertures of size $2\,{\rm mm} \leq D \leq 70\,{\rm mm}$ were used, spanning scales from recreational hourglasses to small industrial chutes. 
About half of the openings are machined from stainless steel, and the others are laser-cut from PMMA to the desired sizes; aperture thickness was also varied. 
Details of each opening size, its corresponding material, and thickness are shown in Table \ref{tab:aperture_details}. 
None of these changes affected the measured flux trends, underscoring that the measurements are robust across different boundary materials and thicknesses. 
This robustness is consistent with our overall framework: confinement effects are governed primarily by packing near boundaries, rather than by specific material or its thickness. 
Images of grains passing through selected apertures are shown in Fig.~\ref{fig:SFig1_Imaging}, which constitute a more exhaustive version of Fig. 1 from the main article.

\begin{table}
\centering
\begin{tabular}{||c|c|c|c||}
\hline \hline
$d$ [mm] & $d_{\rm std. \, dev}$ [mm]& $\rho_g$ [kg/m$^3$] & $\phi_{\rm cyl}$ [adim] \\
\hline \hline
0.26 & 0.03 & 2520 & 0.61 \\
\hline
0.55 & 0.06 & 2500 & 0.60 \\
\hline
1.15 & 0.12 & 2540 & 0.60 \\
\hline
2.10 & 0.21 & 2560 & 0.61 \\
\hline
4.25 & 0.40 & 2510 & 0.60 \\
\hline \hline
\end{tabular}
\caption{Physical properties of the spherical glass particles used in the experiments.}
\label{tab:particle_properties}
\end{table}

\begin{table}
\centering
\begin{tabular}{||c||c|c||}
\hline \hline
\textbf{$D$ [mm]} & \textbf{Material} & \textbf{Thickness [mm]} \\
\hline \hline
2.0  & PMMA            & 2.8 \\
2.5  & Stainless Steel & 1.6 \\
3.3  & Stainless Steel & 1.5 \\
4.0  & Stainless Steel & 1.4 \\
4.7  & Stainless Steel & 1.5 \\
5.0  & PMMA            & 6.0 \\
5.6  & Stainless Steel & 1.5 \\
6.7  & 3D Printing Resin &  5.5\\
7.2  & Stainless Steel & 1.6 \\
8.0  & Stainless Steel & 6.2 \\
8.7  & Stainless Steel & 1.5 \\
9.5  & Stainless Steel & 2.7 \\
10.0 & PMMA            & 6.0 \\
11.0 & PMMA            & 3.1 \\
11.0 & Stainless Steel & 6.3 \\
11.9 & Stainless Steel & 3.0 \\
12.7 & Stainless Steel & 6.3 \\
14.3 & Stainless Steel & 6.2 \\
16.0 & PMMA            & 3.1 \\
17.5 & Stainless Steel & 6.4 \\
18.5 & PMMA            & 3.1 \\
20.0 & PMMA            & 6.0 \\
21.0 & Stainless Steel & 6.2 \\
22.0 & PMMA            & 3.1 \\
24.0 & Stainless Steel & 6.3 \\
25.5 & PMMA            & 3.1 \\
27.0 & Stainless Steel & 6.2 \\
30.0 & PMMA            & 6.0 \\
33.4 & Stainless Steel & 6.3 \\
35.0   & 3D Printing Resin &  5.5\\
42.0 & PMMA            & 6.0 \\
50.0 & PMMA            & 6.0 \\
60.0 & PMMA            & 6.0 \\
70.0 & PMMA            & 6.0 \\
\hline \hline
\end{tabular}
\caption{Details of apertures used in the experiments.}
\label{tab:aperture_details}
\end{table}

\section{Details on Supplementary Movies}

High-speed videos of granular discharge, corresponding to Fig.~\ref{fig:SFig1_Imaging}, are provided in Supplementary Materials. All recordings were captured at 1000~fps and are played back at 4$\times$ slow motion for clarity. Since the packing in the silo is always $\phi_{\rm cyl} \simeq 0.60$, dilation is clearly visible in the cases where $D \sim d$. When $D \gg d$, the jet of grains displays fluid-like thinning. The width of each video file corresponds to $6$ cm. Specific details for each film are listed below:

\subsection*{Fig. S1 (a): $D = 5$ mm}

\begin{itemize}
    \item a\_5mm\_4mm: $d = 4.25$ mm $\implies D/d = 1.2$ \\ 
    \textbf{[NO FLUX]}
    \item a\_5mm\_2mm: $d = 2.10$ mm $\implies D/d = 2.4$ \\ 
    \textbf{[NO FLUX]}
    \item a\_5mm\_1mm: $d = 1.15$ mm $\implies D/d = 4.3$
    \item a\_5mm\_0\_5mm: $d = 0.55$ mm $\implies D/d = 9.1$
    \item a\_5mm\_0\_3mm: $d = 0.26$ mm $\implies D/d = 19.2$
\end{itemize}

\subsection*{Fig. S1 (b): $D = 10$ mm}

\begin{itemize}
    \item b\_10mm\_4mm: $d = 4.25$ mm $\implies D/d = 2.4$ \\ \textbf{[NO FLUX]}
    \item b\_10mm\_2mm: $d = 2.10$ mm $\implies D/d = 4.8$
    \item b\_10mm\_1mm: $d = 1.15$ mm $\implies D/d = 8.7$
    \item b\_10mm\_0\_5mm: $d = 0.55$ mm $\implies D/d = 18.2$
    \item b\_10mm\_0\_3mm: $d = 0.26$ mm $\implies D/d = 38.5$
\end{itemize}

\subsection*{Fig. S1 (c): $D = 20$ mm}

\begin{itemize}
    \item c\_20mm\_4mm: $d = 4.25$ mm $\implies D/d = 4.7$
    \item c\_20mm\_2mm: $d = 2.10$ mm $\implies D/d = 9.5$
    \item c\_20mm\_1mm: $d = 1.15$ mm $\implies D/d = 17.4$
    \item c\_20mm\_0\_5mm: $d = 0.55$ mm $\implies D/d = 36.4$
    \item c\_20mm\_0\_3mm: $d = 0.26$ mm $\implies D/d = 76.9$
\end{itemize}

\subsection*{Fig. S1 (d): $D = 42$ mm}

\begin{itemize}
    \item d\_42mm\_4mm: $d = 4.25$ mm $\implies D/d = 9.9$
    \item d\_42mm\_2mm: $d = 2.10$ mm $\implies D/d = 20$
    \item d\_42mm\_1mm: $d = 1.15$ mm $\implies D/d = 36.5$
    \item d\_42mm\_0\_5mm: $d = 0.55$ mm $\implies D/d = 76.4$
    \item d\_42mm\_0\_3mm: $d = 0.26$ mm $\implies D/d = 161.5$
\end{itemize}

\section{Details on DEM Simulations}

\begin{figure*}
\centering
\includegraphics[width =0.95\textwidth]{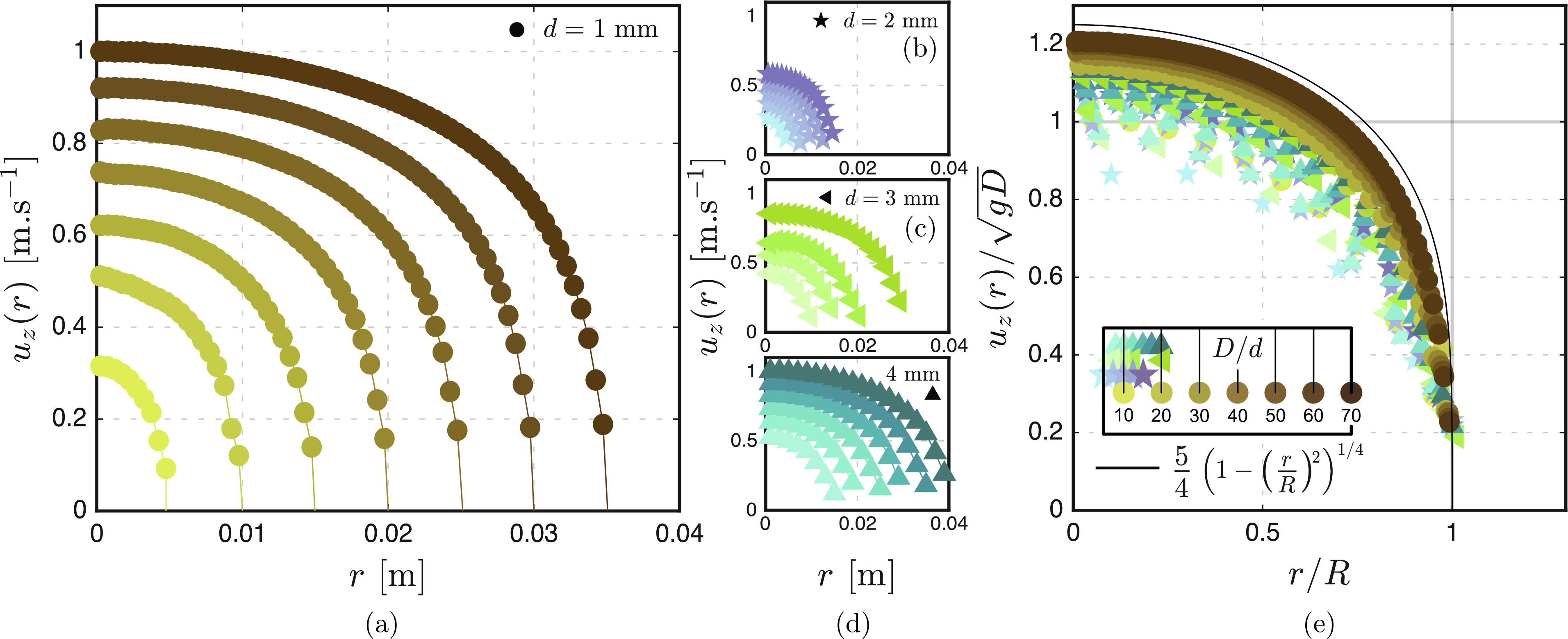}
\caption{Velocity profiles obtained from DEM simulations. The data is averaged in time, as well as along the $\theta$ coordinate. (a) Velocity  profiles for $d = 1$ mm for opening sizes $D  =10, \, 20, \, 30, \, 40,\, 50,\, 60$ and $70$ mm. (b)-(d) Velocity profiles for a few aperture sizes with $d = 2,\, 3$ and $4$ mm particles respectively. Marker shapes indicate grain sizes. (e) Velocity profiles normalized by the free-fall velocity scale $\sqrt{gD}$. 
The radial coordinate is normalized by the radius of the aperture $R=D/2$. For large $D/d$, the profiles collapse on a master curve. A model using free fall is proposed, with $u_z (r) = 5/4 \, \sqrt{gD} \, (1 - (2r/D)^2)^{1/4}$ and shown with the data in a solid line.}
\label{fig:SFig2_Velocity}
\end{figure*}

We also perform Discrete Element Method (DEM) simulations replicating the experimental conditions using the LIGGGHTS software package \cite{liggghts}. A flat-bottomed cylinder (diameter $D_\text{cyl} = 20 \,\text{cm}$) is filled with $7\,\text{kg}$ of spherical particles of diameters, $d=1, \, 2,\, 3\, $ and $4\,\text{mm}$ and density $\rho_g=2500\, \text{kg}.\text{m}^{-3}$. The Hertz-Mindlin model with tangential history force and Constant Directional Torque (CDT) rolling friction model are employed for grain-grain and grain-wall contacts. The micromechanical parameters are set to generally accepted values for efficient simulation of glass beads of $d=\mathcal{O}(1\text{mm})$: coefficient of restitution, $e=0.5$, Young's Modulus, $E=1\, \text{MPa}$, Poisson ratio, $\nu=0.2$, sliding friction coefficient, $\mu_s=0.5$, and coefficient of rolling friction, $\mu_R=0.01$ \cite{dumont_micro_params_2020}. A single initial packing is generated for each particle diameter. A circular aperture of diameter $10\leq D \leq 70\, \text{mm}$ is created at the base, and the cylindrical silo is drained. Altogether, this gives us a large range of $D/d$, spanning the entire range of behavior well into the dense free-fall phase observed in experiments. \smallskip

While draining, Lagrangian data for particle positions, velocities, collision forces, and the number of contacts are recorded at a frequency of 20 Hz. Like in the main article, to avoid transients, we report statistics over the $N_\text{T}$ data files which fall within the time interval, $\Delta T = [T_1, T_2]$ corresponding to when the $0.5$ kg and $2.5$ kg of grains have exited the silo, respectively. To compute integrated quantities such as the volume fraction, we define a thin cylindrical control volume, $\Omega$, of diameter $D_\text{sample}=D$, \textit{i.e.} the aperture, and height $h$, positioned so that its center surface coincides with the base of the cylinder, $y=0$. The volume fraction is computed within this ``puck''-like control volume, by further considering the limit where $h \rightarrow 0$. For our simulations, this corresponds to $h = d/10$. This packing is found by integrating,
\begin{equation}
    \langle \phi\rangle = \frac{1}{V_\Omega \Delta T}\int_{T_1}^{T_2}\oint_{\partial\Omega}\phi(\mathbf{x}, t)\, \dd V\, \dd t
\end{equation}
where we introduce an Eulerian framework with $\phi(r, \theta, z, t)$, the instantaneous volume fraction field. The mass flux can be similarly constructed,
\begin{equation}
    \langle Q \rangle = \frac{1}{V_\Omega \Delta T}\int_{T_1}^{T_2}\oint_{\partial\Omega} \rho_p \phi(\mathbf{x}, t)v_p(\mathbf{x}, t)\, \dd V\, \dd t
\end{equation}
where $v_p(\mathbf{x}, t)$ is the Eulerian particle velocity field. Lastly, we compute the phase-averaged particle velocity by computing,
\begin{equation}
    \langle v\rangle =  \frac{1}{\Delta T}\int_{T_1}^{T_2}\frac{\oint_{\partial\Omega}\phi(\mathbf{x, t})v(\mathbf{x}, t)\, \dd V}{\oint_{\partial\Omega}\phi(\mathbf{x}, t)\, \dd V}\, \dd t.
\end{equation}
In the puck region, we employ the level-set approach of \citet{kempe_level_set_2013} to construct the discrete Eulerian fields $\phi_{i, j, n }= \phi(\mathbf{x}_{i, j}, t_n)$ and $v_{i, j, n} = v(\mathbf{x}_{i, j}, t_n)$ on a cubic Cartesian mesh with grid spacing $\Delta x=d/10$. Note that the sums over the vertical direction are omitted since the region is only a single grid cell tall.
The discretized integrals are then computed over the Eulerian fields,
\begin{equation}\label{eq:discrete_phi}
    \langle \phi \rangle = \frac{\Delta x^3}{V_\Omega N_\text{T}}\sum_n^{N_\text{T}}\sum_{i, j\in\mathcal{I}}\phi_{i, j, n}
\end{equation}
where $\mathbf{x}_{i, j} = (x_i, y_j, z=0)$ and the spatial sums over $i$ and $j$ are over the subset that satisfies $\mathcal{I} = \left\{i, j | \sqrt{x_i^2 + y_j^2} < D\right\}$.
The mass flux and phase-averaged particle velocity follow,
\begin{equation}\label{eq:discrete_Q}
    \langle Q \rangle = \frac{\Delta x^3}{V_\Omega N_\text{T}}\sum_{n=1}^{N_\text{T}}\sum_{i, k\in\mathcal{I}}\phi_{i, j, n}v_{i, j, n}
\end{equation}
and,
\begin{equation}\label{eq:discrete_v}
    \langle v_p \rangle = \frac{1}{ N_\text{T}}\sum_{n=1}^{N_\text{T}}\left(\sum_{i, j\in\mathcal{I}}\phi_{i, j, n}v_{i, j, n}{\sum_{i, j\in\mathcal{I}}\phi_{i, j, n}}\right)
\end{equation}
The radial profiles of the phase-averaged velocity and volume fraction are reported in Figs. \ref{fig:SFig2_Velocity} and \ref{fig:SFig5_Packing} respectively. We construct a radial histogram with bin centers $r_m = m\,d/2 \quad (m=0,1,\dots, 2D/d)$, with spacing $\Delta r = d/2$. The sums \eqref{eq:discrete_phi} and \eqref{eq:discrete_v} are taken within each bin:
\begin{equation}
\langle \phi \rangle(r_m) = \frac{1}{N_\text{T}N_m} \sum_{n=1}^{N_\text{T}}\sum_{i,j \in \mathcal{I}_m} \phi_{i, j, n},
\end{equation}
\begin{equation}
\langle v \rangle(r_m) = \frac{1}{N_\text{T}} \sum_{n=1}^{N_\text{T}}\left(\frac{\sum_{i,j \in \mathcal{I}_m} v_{i, j, n}\phi_{i, j, n}}{\sum_{i,j \in \mathcal{I}_m} \phi_{i, j, n}}\right)
\end{equation}
where, $r_m$ is the center of the $m$-th radial bin, $\mathcal{I}_m = \left\{ i,j \;\middle|\; r_m - (\Delta r/2) < \sqrt{x_i^2+y_j^2} \leq r_m + (\Delta r/2) \right\}$ is the set of grid indices in the $m$-th radial bin, and $N_m$ is the number of grid cells in that bin. The radial bins and resultant profiles are shown in the inset of Fig. \ref{fig:SFig4_Simulations_Phi} for $D/d = 20$ while the mean values over the entire control volume are presented in Fig. \ref{fig:SFig4_Simulations_Phi}.


\section{Velocity of particles at the aperture}

\begin{figure}
\centering
\includegraphics[width =0.95\columnwidth]{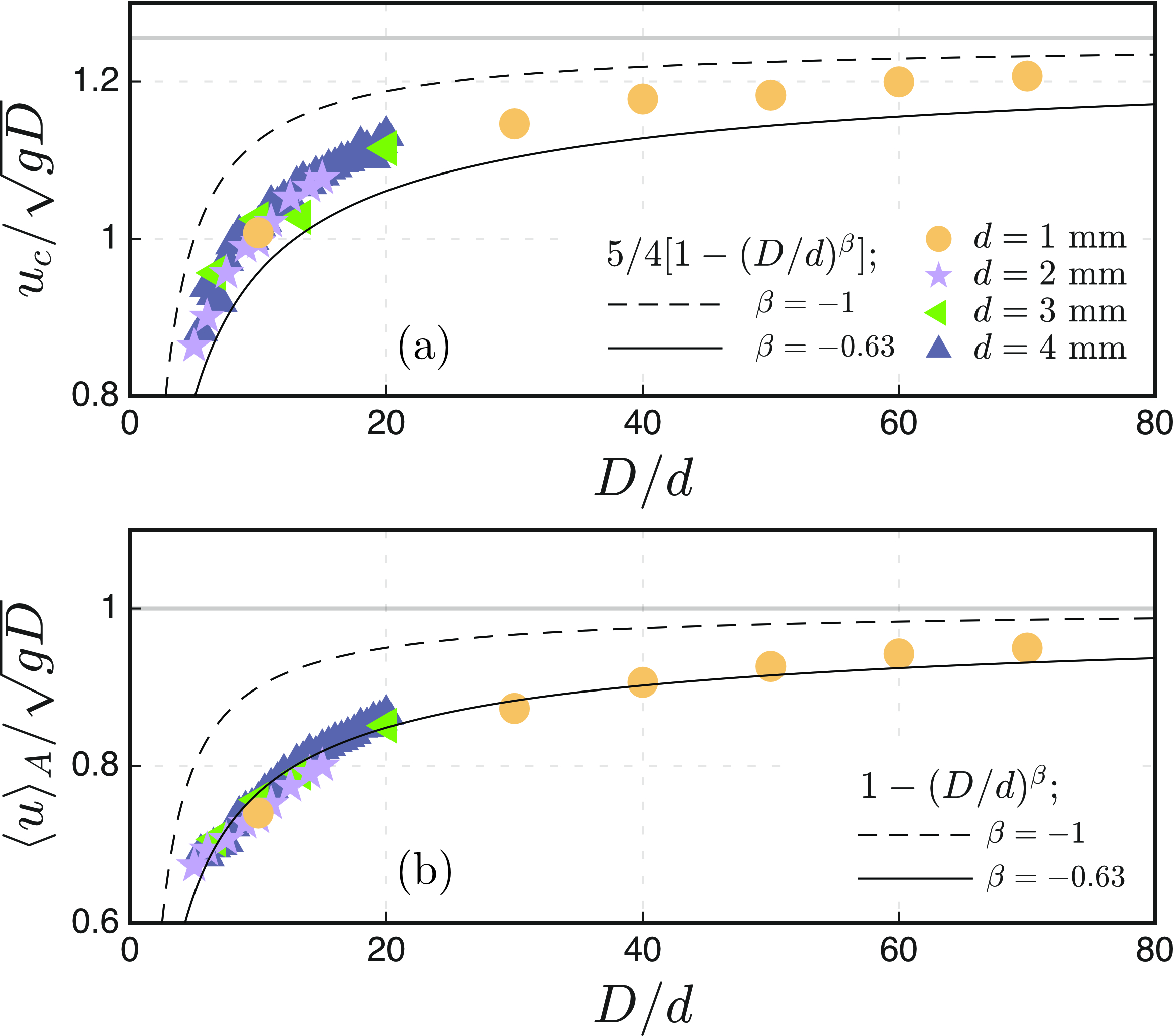}
\caption{(a): Centerline velocity $u_c$ obtained for the simulations as a function of $D/d$. For $D \gg d$, the centerline velocity measured approaches $5 \sqrt{gD}/4$, as predicted by Eq. \eqref{eq:velocity_toy}. (b) Mean velocity $\langle u_z\rangle_A / \sqrt{gD}$ as a function of $D/d$.  Interpolations are shown for the transient behavior, $\langle u_{z} \rangle_A/\sqrt{gD} = 1 - (D/d)^{\beta}$ are shown, with $\beta = -1$ in a dashed line and $\beta = -0.63$ in a solid line.}
\label{fig:SFig3_Centerline}
\end{figure}

In the article, we proposed $\sqrt{gD}$ as the primary velocity scale for the mean downward velocity of the particles at the aperture, $\langle u_z\rangle_A$. This follows from a simple kinematic argument: a particle in free fall from rest over a distance equal to the aperture radius, $L=D/2$, acquires a velocity $\sqrt{2gL}=\sqrt{gD}$. Thus, the relevant length scale in the velocity scale is the aperture radius. \smallskip

Figure~\ref{fig:SFig2_Velocity} shows some examples of velocity profiles from DEM for a range of $D$ and 
$d$. In Fig. \ref{fig:SFig2_Velocity} (a), we show the measured profiles for $D = 10, \, 20, \, 30,\, 40, \,50, \,60$ and $70$ mm in dimensional terms. Across regimes, the curves are very similar. Additional profiles are shown in Figs. \ref{fig:SFig2_Velocity} (b)-(d) using grains of diameter $d = 2, \,3$ and $4$ mm, respectively, for a few opening sizes.\smallskip

When rescaled by the aperture radius $R=D/2$, and the velocity $\sqrt{gD} = \sqrt{2gR}$ (Fig.~\ref{fig:SFig2_Velocity} (e)), the profiles collapse at large $D/d$. The centerline velocity ($u_c$) is larger than $\sqrt{gD}$, consistent with prior studies \cite{janda2012flow}. Averaging over the aperture must yield $\langle u_z \rangle_A = \sqrt{gD}$ once the boundary condition $u_z (r = R) \rightarrow 0$ at the edge is enforced, and therefore $u_c > \sqrt{gD}$. \smallskip

To rationalize these profiles, we adapt a free-fall model inspired by \citet{janda2012flow}. We consider a hemispherical dome of base diameter $D$ above the aperture, from which particles fall from rest. We assume that the velocities of the grains prior to reaching this region are negligible. Although no such dome physically exists \cite{rubio2015disentangling}, it provides a useful kinematic reference. The free-fall height is $h(r) = (R^2 - r^2)^{1/2}$, giving \begin{equation}
    u_z(r) = B \sqrt{2g\, h(r)} = B \sqrt{2gR} \,\,\, f(r)
    \label{eq:S_velocity_general}
\end{equation} where B is a scale factor introduced to ensure the mean over the aperture. Details of the shape are captured through the function $f(r) = \left[1 - (r/R)^2\right]^{1/4}$. Enforcing $\langle u_z \rangle_A = \sqrt{2gR}$, leads to $B = 1/\langle f(r) \rangle_A= 5/4$. Altogether, a model for the velocity profile based only on free-fall is:\begin{equation}
    u_{z,\, \mathrm{ff}} (r) = \frac{5}{4} \, \sqrt{2gR} \, \left(1 - \Big(\frac{r}{R}\Big)^2\right)^{1/4}. 
    \label{eq:velocity_toy}
\end{equation} Simulation profiles (Fig.~\ref{fig:SFig2_Velocity}(e)) indeed approach this free-fall prediction at large $D/d$. This model predicts a centerline velocity $u_c = B \, \sqrt{gD} = (5/4)\sqrt{gD}$ in the limit $D \gg d$. In Fig.~\ref{fig:SFig3_Centerline} (a), we extract $u_{\rm c}$ from the central radial bin of each profile. We also report in Fig.~\ref{fig:SFig3_Centerline} (b) the corresponding mean velocities $\langle u_z \rangle_A$ for the range of $D/d$. Fig. \ref{fig:SFig3_Centerline} illustrates that both $u_c$ and $\langle u_z \rangle_A$ indeed approach the free-fall case for $D/d \gg 1$. \smallskip

For the transition regime, a fit of the form is obtained \begin{equation}
\frac{\langle u_z \rangle_A}{\sqrt{gD}} =  1 - \Big(\frac{D}{d}\Big)^{\beta},
\end{equation} is found to fit the data for the mean velocity [Fig.~\ref{fig:SFig3_Centerline} (b)]. This is analogously cast in terms of $B$ to predict the radial profiles in terms of \eqref{eq:S_velocity_general}:\begin{equation}
B \simeq \frac{5}{4}\Big[ 1 - \Big(\frac{D}{d}\Big)^{\beta}\Big].
\end{equation} The simple choice of power $\beta = -1$ recovers a Beverloo-style algebraic correction (dashed line) \cite{beverloo1961flow, nedderman1982flow, mankoc2007flow}. The best fitting power for the mean velocity ($\beta = -0.63$, RMSE: 0.01) is shown in a solid line in both. The relevant ingredients are captured here without adjustable parameters, showing that gravity and the aperture radius set the fundamental velocity scale. Altogether, we have a profile across $D/d$, written purely in terms of radii ($r_p = d/2$): \begin{equation}
    u_{z} (r) = \frac{5}{4}\Big[1 - \Big(\frac{R}{r_p}\Big)^{\beta} \Big]\sqrt{2gR} \, \left(1 - \Big(\frac{r}{R}\Big)^2\right)^{1/4}. 
    \label{eq:eq_vel_full_final}
\end{equation} Model predictions are shown alongside the measured profiles in Fig. \ref{fig:SFig6_Prediction} (b) for a few simulations, spanning dilute to dense discharge.

In summary, in the limit $D \gg d$, the mean velocity of the particles is $\langle u_z \rangle_A \simeq \sqrt{gD}$, with the centerline $u_c = 5/4 \, \sqrt{gD}$. A small correction through $(D/d)^{\beta}$, with $\beta \approx -1$ describes the transition towards this asymptote. Thus, gravity and a length scale associated with the aperture control the velocity. \smallskip

\begin{figure}
\centering
\includegraphics[width =\columnwidth]{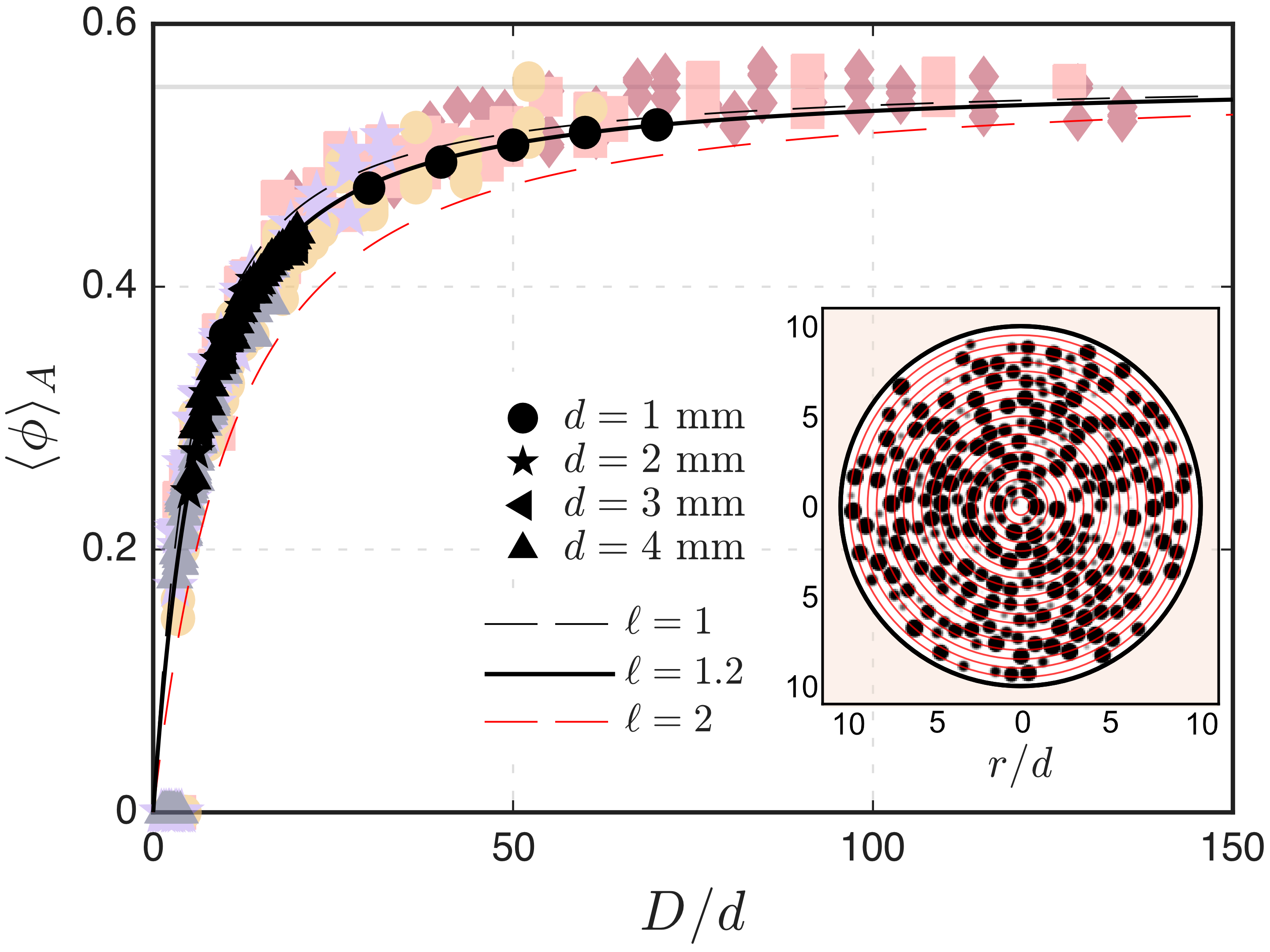}
\caption{Evolution of the mean packing at the aperture during steady discharge $\langle \phi \rangle_A$ as a function of $D/d$. Simulation results are shown in black markers, with the shape indicating the grain size.  Experimental measurements of flux are used to estimate $\langle \phi \rangle_A$, using $\beta = -1$. The Knudsen-like exponential predictions for the mean flux over the entire aperture are shown with mean free path, $\ell = 1d$ (dashed black), $2d$ (dashed red), and $1.2d$ (solid line - best fit for simulations). Inset: Example of the aperture at one time-step for a simulation with $D=20$ mm and $d = 1$ mm. Radial bins have a width $d/2$.}
\label{fig:SFig4_Simulations_Phi}
\end{figure}

\section{Packing of particles at the aperture}

\begin{figure*}
\centering
\includegraphics[width =0.95\textwidth]{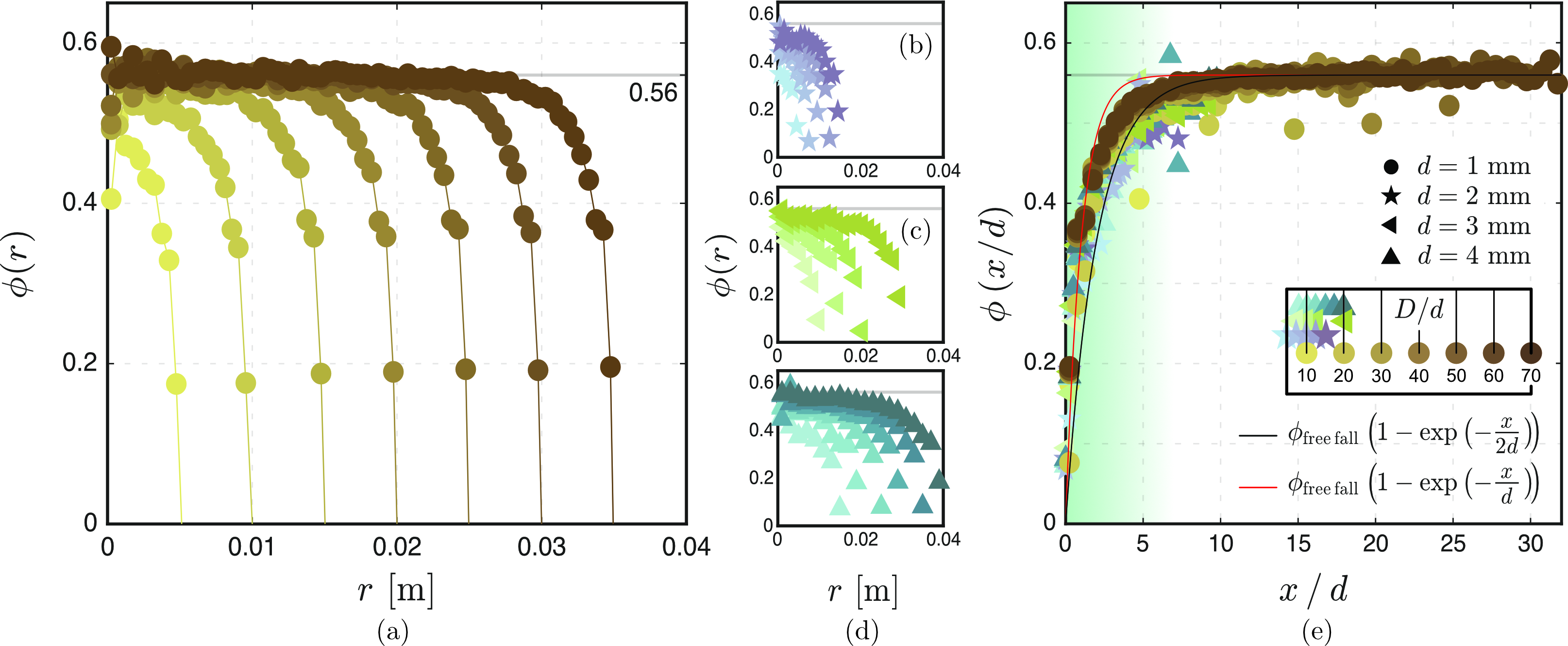}
\caption{Packing profiles obtained from DEM simulations. Data presented here is averaged in time, as well as the angular coordinate. (a) Profiles for $d = 1$ mm for opening sizes $D  =10, \, 20, \, 30, \, 40,\, 50,\, 60$ and $70$ mm. (b) - (d) Profiles with $d = 2,\, 3$ and $4$ mm grains respectively for a few aperture sizes. (e) Packing profiles replotted in wall-normal coordinates normalized by particle size, $x/d =(D/2 - r)/d$. A relaxation equation is shown in solid lines, $\phi = \phi_{\rm ff} \left(1 - {\rm exp}(-x/\ell)\right)$, using $\phi_{x \gg d} = \phi_{\rm ff} = 0.56$, and mean free path $\ell = 1d$ (red) and $\ell = 2d$ (black).}
\label{fig:SFig5_Packing}
\end{figure*}

Figure~\ref{fig:SFig4_Simulations_Phi} summarizes the aperture-averaged packing fraction $\langle \phi \rangle_A$ from simulations across a large range of $D/d$. Experimental flux data were converted to effective packing fractions using the conservation of mass \begin{equation}
 \langle \phi \rangle_A  = \frac{Q}{A \, \rho_g \sqrt{gD} (1 - (D/d)^{-1})},  
\end{equation} \textit{i.e.} $\beta = -1$. While the experimental values lie slightly above the simulations, both converge toward the free-fall limit $\phi_{\rm ff} \approx 0.55$ as $D/d \rightarrow \infty$ and show a very comparable functional form. \smallskip

The profiles of the packing are reported in Fig.~\ref{fig:SFig5_Packing}. Figs.~\ref{fig:SFig5_Packing}(a) - (d) show the range from dilute to dense discharges in experiments using $d = 1,\,2,\,3$ and $4$ mm grains for a range of $D/d$. For $D\gg d$, the central region shows a consistent plug with $\phi \simeq 0.56$, in agreement with prior measurements \cite{prado2013incompressible,onoda1990random}, and our estimates with a free fall velocity. At the rim, the packing decreases very rapidly. \smallskip

Expressed in wall-normal coordinates $x = (D/2) - r$ and normalized by grain size $d$, all simulation data collapse onto a single curve (Fig.~\ref{fig:SFig5_Packing} (e)). The transient region extends over $\lambda/2 \simeq 7-8d$ (shaded region), consistent with the scale $nd$ extracted from the flux ratio ($n \approx 10-15$). Thus, when $D/2 \gg \lambda/2$, the plug region at $\phi_{\rm ff}$ dominates and the flux approaches the free-fall limit, $Q_{\rm ff}$. \smallskip

\begin{figure*}
\centering
\includegraphics[width =0.95\textwidth]{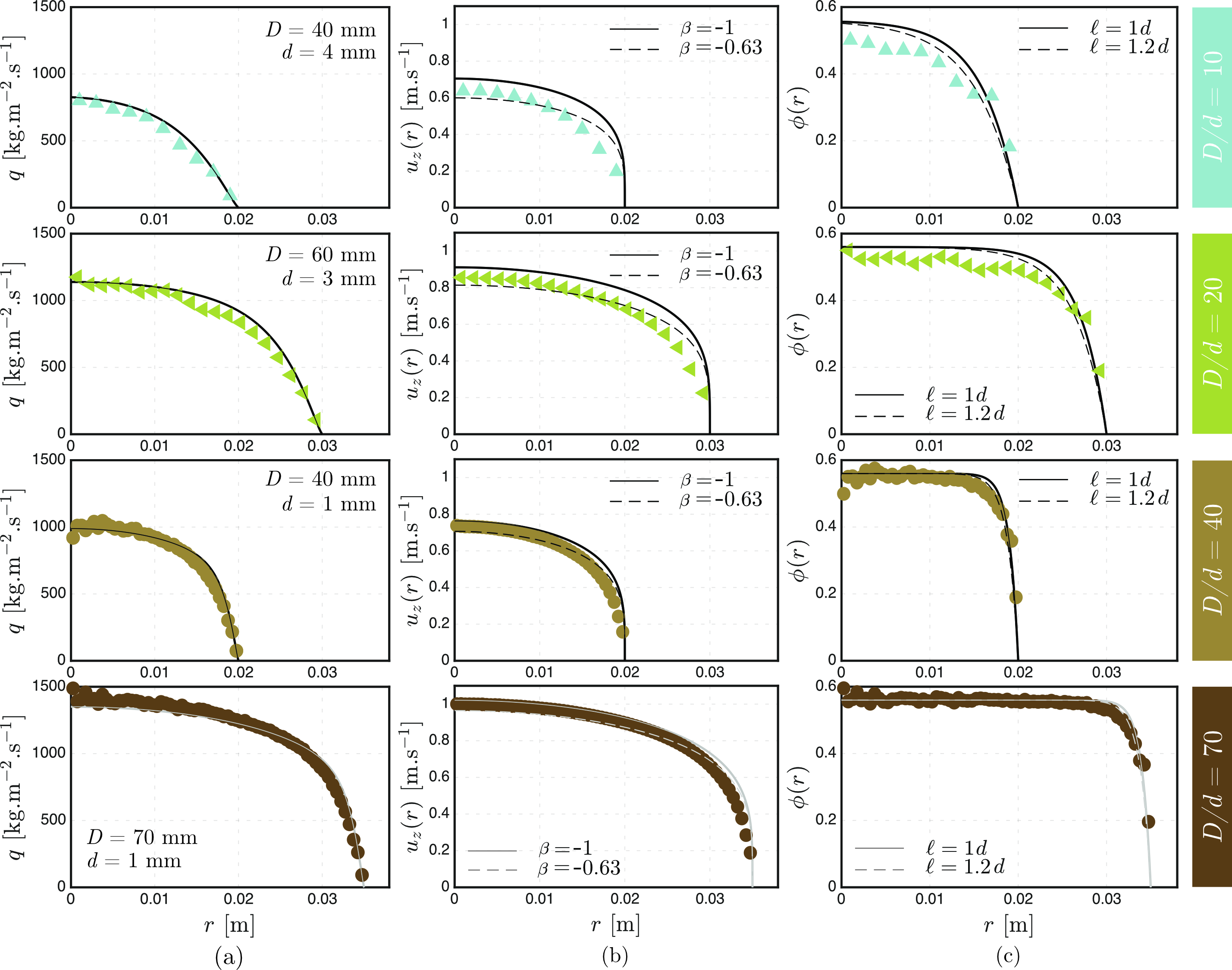}
\caption{Comparison of the model profiles and simulation data for (a) the local mass flux density $q (r)$, (b) the velocity $u_z (r)$, and (c) the packing $\phi (r)$ at the aperture. Top to bottom: $D/d = 10, \, 20,\, 40$ and $70$. (a) The best fitting values of $\beta = -0.63$ and $\ell= 1.2 \, d$ together predict the overall flux density. Models are shown for both (b) $u_z(r)$ Eq.\eqref{eq:eq_vel_full_final} [Dashed line: $\beta = -0.63$; Solid line: $\beta = -1$] and (c) $\phi(r)$ Eq. \eqref{eq:eq_phi_full_final} [Dashed line: $\ell = 1.2 \, d$; Solid line: $\ell = -1$].}
\label{fig:SFig6_Prediction}
\end{figure*}

The relaxation near the boundary is modeled as a structural analogue of a Knudsen layer. A kinetic mean free path for hard spheres is described as: $\ell_k = (\sqrt{2}\,\pi \,d^2 \,N)^{-1}$, where $n$ is the number density \cite{sone2007molecular}. For a dense system, using $N = \phi/v_g$ where $v_g = \pi\, d^3/6$ is the volume of a grain and $\phi \approx 0.56$, we get $\ell_k \approx 0.2 \, d$. 
The structural mean free path $\ell$ describes the distance over which packing recovers from a boundary, \textit{i.e.}, a measure of the length over which structure persists in units of particle size. Borrowing the exponential form from gas kinetics \cite{sone2007molecular}, we have \begin{equation}
    \phi(x) = \phi_{\rm ff} \left[1 - {\rm exp} \left({-\frac{x}{\ell}}\right)\right].
    \label{eq:phi_x}
\end{equation} with $\ell = \mathcal{O}(d)$ for cohesionless grains, as the microscopic physical scale in discharge. Simulation data shows that $\ell$ between $1\,d$ and $2\, d$ captures the data [Fig.~\ref{fig:SFig5_Packing} (e)]. Recasting into radial form, \begin{equation}
    \phi(r) = \phi_{\rm ff} \left[1 - {\rm exp} \left({\frac{r-R}{\ell}}\right)\right].
    \label{eq:eq_phi_full_final}
\end{equation} This model is shown alongside the measured profiles from simulations in Fig.~\ref{fig:SFig6_Prediction} (c), spanning dilute to dense discharge. Averaging this distribution over the entire aperture gives
\begin{equation}
    \frac{\langle \phi \rangle_A}{\phi_{\rm ff}} = 1 - \frac{4 \ell}{D} + \frac{8\ell^2}{D^2}\left(1 - {\rm exp}\left(-\frac{D}{2\ell}\right)\right).
\end{equation} Fits to simulation data yield $\ell \simeq 1.2\, d$, while experiments are also well described by $\ell \approx 1\, d$ (Fig.~\ref{fig:SFig4_Simulations_Phi}). \smallskip

The exponential recovery of packing near the wall \eqref{eq:phi_x} also reflects the excluded-volume relaxation observed in dense granular systems. In static sphere packings, \citet{benenati1962void} measured a similar void-fraction profile, showing that particle centers are geometrically forbidden within one grain diameter of the boundary and that the packing fraction approaches its bulk value exponentially over a few particle diameters. In kinetic-theory treatments of rapid granular flows, \citet{jenkins1983theory} introduced the same idea as a boundary correction arising from excluded-volume and reduced collision frequency, yielding a first-order relaxation equation \begin{equation}
\frac{\dd \phi}{\dd x} = \frac{\phi_{\rm ff} - \phi(x)}{\ell},
\end{equation} and since only point contacts are present at the wall, the packing must vanish, $\phi(0) =0$. The solution to this equation is precisely \eqref{eq:phi_x}. More globally, \citet{onoda1990random} showed that the loss of mechanical stability at the random-loose-packing threshold corresponds to the same geometric exclusion operating throughout the bulk. Together, these studies establish the exponential recovery form as a generic structural consequence of excluded volume, with the present mean-free-path length $\ell \simeq 1-2 \, d$ consistent with their reported scales.

In summary, the granular packing adjusts over a structural mean free path of order one grain diameter, creating a boundary layer of thickness $\sim 7\, d-8 \, d$. This exponential recovery toward $\phi_{\rm ff}$ explains the confinement dependence of the flux, while the velocity remains essentially at the free-fall scale $\sqrt{gD}$.

\section{Combining distributions to flux}

We are able to provide an expression of the flux from the two individual distributions found using the simulations (S.IV. and S.V.). The dimensionless flux number $F$, from the main article is: \begin{equation}
    F = \frac{\langle u_z \phi\rangle_A}{\sqrt{gD}\, \phi_{\rm ff}} = \Big(\frac{\langle u_z\rangle_A}{\sqrt{gD}}\Big)\Big(\frac{\langle \phi\rangle_A}{\phi_{\rm ff}}\Big).
\end{equation} To confirm the decomposition of $\langle u_z\, \phi\rangle_A$ into $\langle u_z\rangle_A\langle \phi\rangle_A$ is valid, we compute the effective error through the covariance: $
|\langle u_z \, \phi \rangle_A - \langle u_z \rangle_A\langle\phi \rangle_A|/ \langle  u_z \, \phi\rangle_A$. This quantity rapidly decreases, suggesting that such a decomposition is justified, in particular when $D \gg d$. \smallskip

\begin{figure*}
\centering
\includegraphics[width =0.9\textwidth]{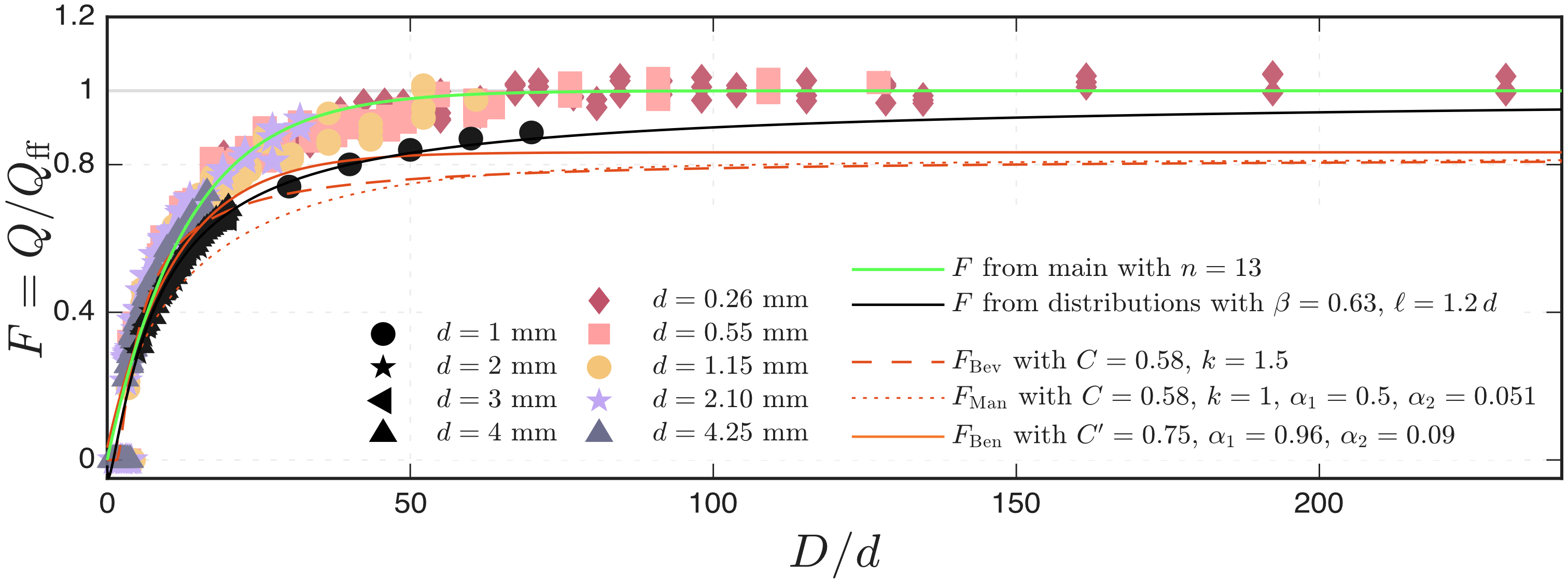}
\caption{Evolution of the dimensionless flux number $F = Q/Q_{\rm ff}$ as a function of the effective aperture size $D/d$. For both experiments (color) and simulations (black), the data collapse onto a single trend. $F = 1 - {\rm exp} (-D/nd)$ with n=13 shown in a green solid line Eq. \eqref{eq:eq_flux_exp}. Eq. \eqref{eq:eq_flux_full_micro} with best fitting parameters $\beta = 0.63$ and $\ell = 1.2\, d$ for velocity and packing confinement effects, respectively, from simulations (solid black line). Additional comparisons with \citet{beverloo1961flow} (dashed), \citet{mankoc2007flow} (dotted) and \citet{benyamine2014discharge} (solid orange line) are also shown.}
\label{fig:SFig7_Exp_Sim}
\end{figure*}

We describe the flux from the distributions of velocity and packing. We introduce two parameters: (i) a power-law exponent for the velocity (best fit: $\beta=-0.63$); and (ii) a structural mean free path $\ell$, which sets the range over which packing is modified by the boundary. When $D$ and $\ell$ are comparable, dilation near boundaries is non-negligible (best fit: $\ell=1.2\, d$). Combining the two contributions yields \begin{equation}
    F = \Big[ 1 - \Big(\frac{D}{d}\Big)^{\beta}\Big] \Big[ 1 - 4\Big(\frac{D}{\ell}\Big)^{-1} + 8\Big(\frac{D}{\ell}\Big)^{-2}\left(1 - e^{-\frac{D}{2\ell}}\right)\Big].
    \label{eq:eq_flux_full_micro}
\end{equation} Using the best-fit values of $\beta$ and $\ell$ gives excellent agreement with simulations as shown in Fig.~\ref{fig:SFig7_Exp_Sim} (black markers: simulation $Q/Q_{\rm ff}$ with $\phi_{\rm ff} =0.56$); dashed line: \eqref{eq:eq_flux_full_micro}). For comparison, results from experiments are also plotted, and saturate over a very comparable $D/d$ to the simulations. \smallskip

Since $\beta \approx -1$, and $\ell \approx 1\, d$ for spheres falling through a circular aperture, this expression \eqref{eq:eq_flux_full_micro} can be interpreted simply in terms of $D/d$. At small $D/d$, the exponential term dominates, leading to a quick decrease of $F$. For larger $D/d$, the approach to the asymptote is slower, and the $\mathcal{O} (D/d)^{-1}$ term dominates, as the expression approaches 1. Defining a Knudsen number $\mathrm{Kn} = (D/d)^{-1}$ like the main article, the Knudsen-layer analogy for gases can be made explicit \cite{sone2007molecular}, as a combination of exponential and power-law components for the transition between particular and fluid-like regimes. When $\mathrm{Kn} \lesssim 10^{-2}$, packing has reached its largest value without maintaining contacts/loads, and $F \simeq 1$ resembling a continuum. \smallskip

This analysis shows that the two-parameter expression [Eq. \eqref{eq:eq_flux_full_micro}] captures the essential impact of confinement: rapid suppression of flux at small $D/d$ from exponential packing relaxation, and a slower algebraic approach to the asymptotic free-fall value at large $D/d$. At the same time, the single-parameter exponential law introduced in the main text [Eq. \eqref{eq:eq_flux_exp}] provides a practical and remarkably accurate description of the data across the accessible range of $D/d$. 
While Eq. \eqref{eq:eq_flux_full_micro} ensures the correct vanishing flux at $D=d$ and encodes the asymptotic algebraic relaxation, the simple exponential \eqref{eq:eq_flux_exp} captures the observed collapse with only one parameter ($n\approx 13$). This makes it especially useful for direct comparison across experiments, simulations, and model systems. \smallskip

\section{Comparison of flux expressions}

In Fig.~\ref{fig:SFig7_Exp_Sim}, we show the dimensionless flux ratio $F$ from experiments (color) and simulations (black markers). The simulations seem to saturate to free fall slower than the experiments, similar to what we noted in Sec. S.V. regarding packing. In the main study, we presented a single-parameter exponential relaxation, which is also shown here (green line),\begin{equation}
    F = 1 - {\rm exp}\Big(\frac{-D}{nd}\Big),
    \label{eq:eq_flux_exp}
\end{equation} with $n = 13$. The value of $n$ is in a reasonable range across experiments and simulations, and comparable with other results on the incompressibility of dense granular jets \cite{prado2013incompressible}. While the primary aim of this study was to define and identify the asymptotic state of granular discharge, such a correction factor allows us to write a one-parameter expression for the flux. In this section, we compare some existing expressions of $Q$ \cite{beverloo1961flow, nedderman1982flow, mankoc2007flow, gella2017role, janda2012flow, benyamine2014discharge} with $Q_{\rm ff}$. First, from \citet{beverloo1961flow}, \begin{equation}
Q_{\rm Bev} = C \rho_b \sqrt{g} \left(D - kd\right)^{5/2}.
\label{eq:eq_Beverloo}
\end{equation} We plot the ratio of this expression with the presented expression for $Q_{\rm ff}$, using $\rho_b = \rho_g \, \phi_b$: \begin{equation}
    F_{\rm Bev} = \frac{Q_{\rm Bev}}{Q_{\rm ff}} = \frac{4 C \phi_b}{\pi \phi_{\rm ff}} \left[1 - k \Big(\frac{d}{D}\Big)\right]^{5/2}
\end{equation} This expression is also plotted on Fig.~\ref{fig:SFig7_Exp_Sim} using the standard values of $C = 0.58$ \cite{nedderman1982flow} and $k = 1.5$ \cite{beverloo1961flow, nedderman1982flow}. For the bulk packing we use the packing measured in the cylinder, $\phi_b = \phi_{\rm cyl} \simeq 0.60$, and if $\phi_{\rm ff} = 0.55$, this expression saturates at $0.806$... for $D \gg d$. \smallskip

Similarly, we also compare with the relation presented by \citet{mankoc2007flow}:
\begin{equation} 
Q_{\rm M, \, 1} = C \, \rho_b \, \sqrt{g}  \, \Big(\frac{D}{d} - 1\Big)^{5/2} \, \left[1 - \alpha_1\,  e^{-\alpha_2(D/d - 1)} \right], \label{eq:eq_Mankoc} 
\end{equation} where they posited $k = 1$, and introduced two new parameters, here labeled $\alpha_1$ and $\alpha_2$. Their data suggested an exponential correction towards Beverloo's law, in particular when $D \sim d$. We should emphasize that an equivalent correction at small $D/d$ was also observed in our experiments and simulations as a consequence of packing modifications of boundaries. The dimensions of $Q_{\rm M, \, 1}$ do not seem to be correct, for the length-scale has been non-dimensionalized. This is likely a typo, and the expression actually contains a $(D - d)^{5/2}$. Indeed in \citet{gella2017role} and \citet{janda2012flow}, this equation was noted (modified) as:
\begin{equation} 
Q_{\rm M, \, 2} = C \, \rho_b \, \sqrt{g}  \, (D - d)^{5/2} \, \left[1 - \alpha_1\,  e^{-\alpha_2(D - d)} \right], \label{eq:eq_Mankoc_Gella} 
\end{equation}
but now a length-scale dimension is given to the exponent. We suggest the following possible interpretation as what the authors perhaps intended (between the two relations):\begin{equation} 
Q_{\rm Man} = C \, \rho_b \, \sqrt{g}  \, (D - d)^{5/2} \, \left[1 - \alpha_1\,  e^{-\alpha_2(D/d - 1)} \right]. \label{eq:eq_Mankoc_corrected} 
\end{equation} A similar comparison to $Q_{\rm ff}$ is made: \begin{equation}
    F_{\rm Man} = \frac{4 C \phi_b}{\pi \phi_{\rm ff}} \left[1 - \Big(\frac{d}{D}\Big)\right]^{5/2} \, \left(1 - \alpha_1\,  e^{-\alpha_2(D/d - 1)} \right)
\end{equation} The saturation value is again set by the same factor containing $C$, $\phi_b$ and $\phi_{\rm ff}$, \textit{i.e.}, it also saturates at $0.806$. In both these relations ($Q_{\rm Bev}$ and $Q_{\rm Man}$) if C were larger, $C \approx 0.71$, these expressions would also saturate $\approx 1$, but the value of $C$ in this geometry is usually noted as $0.55-0.65$ \cite{nedderman1982flow, beverloo1961flow}, likely reflecting the lack of experiments for $D \gg d$. \smallskip

A final comparison is presented with \citet{benyamine2014discharge}. Their relation is:\begin{equation}
    Q_{\rm Ben} = C' \, A \, \rho_g \, \phi_b \left[1 - \alpha_1 \, e^{-\alpha_2(D/d)}\right] \sqrt{gD}.
\end{equation} with three constants, $C$, $\alpha_1$ and $\alpha_2$. A number of similarities are shared in the final expression we find using the dimensionless flux ratio, and the expression presented by \citet{benyamine2014discharge}. Indeed, they also find an exponential correction to the packing. They also use the area of the aperture. A few points of difference are also noted, we argue that in free fall packing asymptotes to a value smaller than the bulk or rest packing ($\phi_{\rm ff} < \phi_{\rm cyl}$). Additionally, our packing profiles are not self-similar in the manner velocity is, and instead display plug-like profiles, showing modifications reminiscent of boundary layers. In comparison with free fall: \begin{equation}
    F_{\rm Ben} = \frac{Q_{\rm Ben}}{Q_{\rm ff}} = \frac{C' \phi_b}{\phi_{\rm ff}} \, \left(1 - \alpha_1\,  e^{-\alpha_2(D/d)} \right).
\end{equation} Using $C' = 0.75$, $\phi_b = \phi_{\rm cyl} \simeq 0.60$, and if $\phi_{\rm ff} = 0.55$, this saturates at $0.818$. Assuming a larger bulk density or making $C' \approx 1$ makes this relation saturate closer to 1. Additionally, since their fits of $\alpha_1 = 0.96 \approx 1$, and $1/\alpha_2 = 11.11$, their expression can otherwise be cast to seem very similar to the one presented in the main article here. 

\bibliographystyle{apsrev4-2}
\bibliography{BEV.bib}